%% file: main.tex
\documentclass[sigconf]{acmart}

\usepackage{acmart-taps}

\AtBeginDocument{%
  }

\copyrightyear{2026}
\acmYear{2026}
\setcopyright{rightsretained}
\setcctype{by}
\acmConference[CHI '26]{Proceedings of the 2026 CHI Conference on Human Factors in Computing Systems}{April 13--17, 2026}{Barcelona, Spain}
\acmBooktitle{Proceedings of the 2026 CHI Conference on Human Factors in Computing Systems (CHI '26), April 13--17, 2026, Barcelona, Spain}
\acmDOI{10.1145/3772318.3790416}
\acmISBN{979-8-4007-2278-3/2026/04}

\input{config}

\begin{document}

\title{\model: Enhancing Creative Storytelling through Multi-Persona Collaborative Improvisation}

\makeatletter
\renewcommand{\@fnsymbol}[1]{\ensuremath{\ifcase#1\or \star \or 1\or 2\or 3\or 4\or \,\textrm{\Letter} \or 5\or\else\@ctrerr\fi}}
\makeatother

\author{Yuxi Ma}
\orcid{0009-0001-2867-9620}
\affiliation{%
  \institution{Institute for Artificial Intelligence, Peking University}
  \city{Beijing}
  \country{China}
}
\email{yxma@stu.pku.edu.cn}
\authornote{Both authors contributed equally to this research.} 
\authornote{Also with the School of Psychological and Cognitive Sciences, Peking University.} 
\authornote{Also with the State Key Lab of General AI.} 
\authornote{Also with the Beijing Key Laboratory of Behavior and Mental Health, Peking University.} 

\author{Yongqian Peng}
\orcid{0009-0003-0564-0787}
\affiliation{%
  \institution{Institute for Artificial Intelligence, Peking University.}
  \city{Beijing}
  \country{China}
}
\email{yqpeng@stu.pku.edu.cn}
\authornotemark[1]
\authornotemark[2]
\authornotemark[3]
\authornote{Also with Yuanpei College, Peking University.} 

\author{Fengyuan Yang}
\orcid{0009-0009-6500-2473}
\affiliation{%
  \institution{Institute for Artificial Intelligence, Peking University}
  \city{Beijing}
  \country{China}
}
\email{fyyang@stu.pku.edu.cn}
\authornotemark[2]
\authornotemark[3]
\authornotemark[5]

\author{Siyu Zha}
\orcid{0009-0003-0292-5162}
\affiliation{%
  \institution{The Future Laboratory, Tsinghua University}
  \city{Beijing}
  \country{China}
}
\email{zhasiyu22@mails.tsinghua.edu.cn}

\author{Chi Zhang}
\orcid{0000-0003-4948-0714}
\affiliation{%
  \institution{School of Intelligence Science and Technology, Peking University}
  \city{Beijing}
  \country{China}
}
\email{chizhang.cz@pku.edu.cn}
\authornotemark[3]

\author{Zixia Jia}
\orcid{0009-0008-6746-0593}
\affiliation{%
  \institution{Beijing Institute for General Artificial Intelligence (BIGAI)}
  \city{Beijing}
  \country{China}
}
\email{jiazixia@bigai.ai}
\authornotemark[3]

\author{Zilong Zheng}
\orcid{0000-0003-1219-5151}
\affiliation{%
  \institution{Beijing Institute for General Artificial Intelligence (BIGAI)}
  \city{Beijing}
  \country{China}
}
\email{zlzheng@bigai.ai}
\authornote{Corresponding authors} 
\authornotemark[3]

\author{Yixin Zhu}
\orcid{0000-0001-7024-1545}
\affiliation{%
  \institution{School of Psychological and Cognitive Sciences, Peking University}
  \city{Beijing}
  \country{China}
}
\email{yixin.zhu@pku.edu.cn}
\authornotemark[6]
\authornotemark[3]
\authornotemark[4]
\authornote{Also with Institute for Artificial Intelligence, Peking University} 
\renewcommand{\shortauthors}{Ma, Peng, \etal.}

\begin{abstract}
Large Language Models show promise for AI-assisted storytelling, yet current tools often generate predictable, unoriginal narratives. To address this limitation, we present \model, a multi-persona co-creative system grounded in Campbell's Blind Variation and Selective Retention theory. \model deploys specialized \ac{ai} personas to generate diverse narrative options (blind variation), while users act as creative directors to select and refine them (selective retention). We designed a controlled study with 50 participants and found that stories co-authored with \model were not only perceived by users as more novel and diverse but were also objectively rated by experts as significantly better across all Torrance Test creativity dimensions: fluency, flexibility, originality, and elaboration. Stories are significantly longer with richer settings and more dialogue. Writing expertise emerged as a moderator: novices benefited more from structured scaffolding. This demonstrates the value of theory-informed co-creative systems and the importance of adapting them to varying user expertise. Project page: \url{https://ppyyqq.github.io/narrativeloom}.
\end{abstract}



\begin{teaserfigure}
    \centering
    \includegraphics[width=\linewidth]{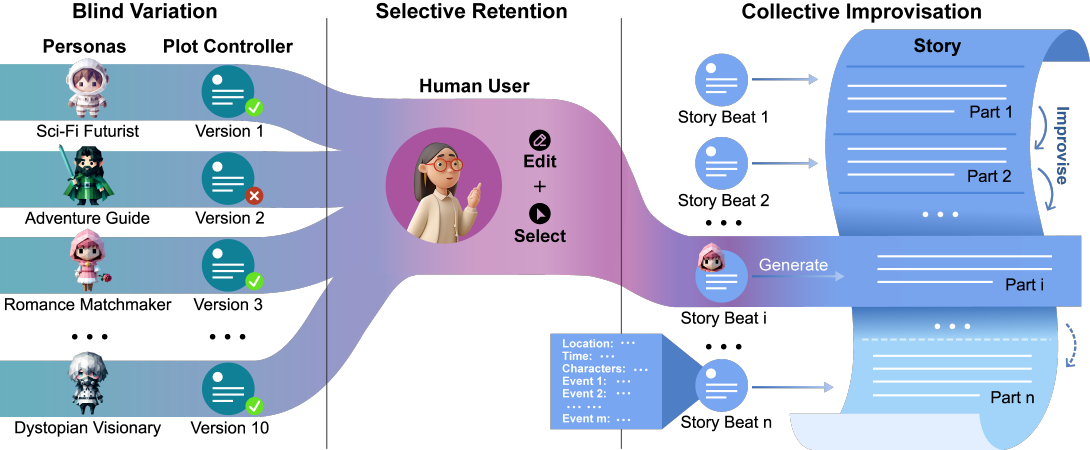}
    \caption{\textbf{The architecture of \model, the multi-persona co-creation system inspired by the \acf{bvsr} theory.} Our computational implementation of \acs{bvsr} operates through three interconnected phases: (i) \BV: Ten specialized storyteller personas independently generate diverse narrative possibilities for each story beat, with a plot controller ensuring baseline coherence through consistency verification; (ii) \SR: The human user evaluates, selects, and optionally edits the most promising beat from the generated alternatives, exercising creative direction; and (iii) \textit{Collective Improvisation}: Selected story beats---containing structured narrative elements (location, time, characters, events)---are sequentially transformed into cohesive narrative segments that progressively build the complete story through collaborative human-\acs{ai} iteration.}
    \Description{A system architecture diagram showing three interconnected phases in a horizontal flow. Phase one shows ten circular persona icons arranged in a semi-circle, each generating story beat alternatives that feed into a plot controller verification module. Phase two depicts a human user figure evaluating and selecting from multiple beat options displayed as cards. Phase three illustrates selected beats being transformed into narrative text segments through an iterative process, with arrows indicating the cyclical nature of beat generation, selection, and expansion that builds the complete story.}
    \label{fig:framework}
\end{teaserfigure}

\maketitle

\section{Introduction}

Storytelling is a fundamental pillar of human culture and a primary medium for sharing knowledge, values, and experiences \citep{lawrence2016our, stone1995images}. Effective narratives, from Aristotle's \textit{Poetics}~\citep{aristotle1942poetics} to modern screenplays, depend on a balance between coherence and surprise. They require ``peripeteia''---unexpected turns that engage audiences not through randomness, but by revealing a deeper, latent logic~\citep{franceschelli2024creativity}. This combination of novelty and appropriateness characterizes human creativity and remains a central focus of computational creativity research~\citep{riedl2006story, gervas2009computational}.

\acp{llm} such as GPT-4~\citep{achiam2023gpt} and Gemini~\citep{team2023gemini} have changed the landscape of computational narrative generation. These models provide higher fluency and contextual understanding than earlier rule-based systems~\citep{alhussain2021automatic, wang2023open}. However, the \ac{hci} community has noted that current \ac{ai} storytelling systems often remain conservative~\citep{chakrabarty2024art, yuan2022wordcraft, mirowski2023co, clark2018creative}. Because their architectures are optimized for next-token prediction, they excel at producing statistically probable continuations but often suppress the surprising deviations essential for creative depth~\citep{chakrabarty2024art, chakrabarty2025can, gervas2009computational, franceschelli2024creativity}. Industry professionals have observed that this limitation~\citep{anderson2024homogenization, chung2022talebrush, clark2018creative, dhillon2024shaping, mirowski2023co} often results in predictable, cliché narratives~\citep{chakrabarty2025can, ippolito2022creative, riedl2006story, yao2019plan}, leading to homogenized ideas.

Our formative study investigating how writers experience these limitations revealed a tension between the desire for creative diversity and the need for narrative control. Current tools often fail to resolve this; they either suggest predictable paths or generate content disconnected from the writer's vision. This highlights a key \ac{hci} challenge: enabling \acs{llm}-based systems to generate meaningful surprises while maintaining user agency.

To address these needs, we applied Campbell’s theory of \acf{bvsr}~\citep{campbell1960blind}. \ac{bvsr} describes a two-phase process: generating unconstrained ideas (``blind variation'') followed by the deliberate curation of promising ones (``selective retention''). This framework serves as a theoretical blueprint for a system that offers a broader set of possibilities while preserving the user's creative authority. Based on this foundation, we developed \model, a system that operationalizes \ac{bvsr} through multi-persona collaborative improvisation~\citep{sawyer2000improvisation, sawyer2009distributed}. To implement ``blind variation,'' the system employs an ensemble of specialized \ac{ai} personas, each providing a unique genre-aware narrative lens to generate diverse story beats. To facilitate ``selective retention,'' users act as creative directors who select, edit, and integrate these beats. Our approach differs from conventional tools by: (i) separating generative and curatorial processes; (ii) achieving variation through specialized personas rather than parametric sampling; and (iii) scaffolding exploration while maintaining user authority.

Our findings demonstrate that \model's \ac{bvsr}-based, multi-persona approach significantly enhances creative outcomes compared to the single-voice chatbot, producing longer, richer stories with more settings and higher dialogue ratios that emphasize ``showing over telling~\citep{howard1993tools}.'' Users perceived \model as providing more diverse narrative possibilities while maintaining high usability, with strategic persona engagement revealing asymmetric transition patterns where certain personas serve as ``initiators'' and others as ``developers.'' Professional writing experts rated stories generated by \model significantly higher across all creativity dimensions---fluency, flexibility, originality, and elaboration---praising its ability to create unexpected narrative turns and psychologically complex characters. Writing experience emerged as a key moderating factor, with novice writers benefiting more from \model's structured scaffolding, suggesting particularly strong support for writers developing narrative intuitions.

This research offers three primary contributions to the fields of \ac{hci} and computational creativity: 
(i) a \textbf{theoretical framework} that operationalizes Campbell's \acf{bvsr} theory within human-AI co-authorship, providing a principled mechanism to balance AI-driven diversity with human executive control; 
(ii) a \textbf{system design} utilizing a multi-persona collective improvisation architecture, which structures creative variation through the functional heterogeneity of specialized personas rather than stochastic parametric sampling; and 
(iii) \textbf{empirical insights} revealing that both writing expertise and creative stages significantly moderate system effectiveness, necessitating adaptive interfaces that accommodate varying skill levels and shifting creative contexts.

\section{Related Work}

\subsection{Human-\acs{ai} Co-creation in Storytelling}

Human-AI co-creation in storytelling has shifted from basic creativity support to collaborative partnerships that redistribute agency between writers and computational systems. Recent \ac{hci} research examines the social and cognitive dynamics of these interactions. For instance, \citet{gero2023social} show how writers navigate the boundary between treating \ac{ai} as a tool \vs a collaborator, while \citet{dhillon2024shaping} find that different scaffolding levels alter both the creative process and narrative quality. Although \acp{llm} provide high fluency, professional writers report concerns regarding creative homogenization and the difficulty of maintaining an authentic voice within systems like Wordcraft \citep{yuan2022wordcraft, ippolito2022creative}.

To address these limitations, the \ac{hci} community has developed interaction paradigms that offer control beyond text generation. \citet{chung2022talebrush} introduced TaleBrush, which uses line-level sketching to shape story arcs, demonstrating that visual modalities can effectively direct creative outcomes. Addressing the need for structural control, \citet{lu2025whatelse} proposed WhatELSE, a system that allows authors to manage narrative abstraction through planning to ensure causal soundness in open-ended plots. Similarly, \citet{mirowski2023co} argue for tools that serve as creative partners rather than generators in screenplay writing, while \citet{chakrabarty2025can} investigated iterative editing interfaces to align human intent with model output.

Despite these innovations, maintaining structural coherence in long-form narratives remains difficult. While hierarchical frameworks use top-down planning to address this \citep{yang2023doc}, \citet{mirowski2023co} argue that rigid structures conflict with the emergent nature of writing, often producing predictable patterns. \citet{chakrabarty2024art} characterize this as a ``false promise of creativity,'' where models optimized for next-token prediction fall into a ``probability trap'' \citep{franceschelli2024creativity}. This results in technically proficient but conservative content that lacks the deviations necessary for compelling storytelling. These tensions between coherence and diversity, and between agency and assistance, indicate a need for approaches that support narrative variety while preserving authorial control.

\subsection{Improvisational Storytelling}

Improvisational storytelling involves the construction of spontaneous narratives without predetermined elements \citep{sawyer2000improvisation,sawyer2009distributed}. This process requires continuous adaptation and generates uncertainty, distinguishing it from hierarchical approaches \citep{hansen2007truth,tanenbaum2008improvisation}. This framework is aligned with the distributed creativity theory \citep{sawyer2014group,sawyer2009distributed}, which posits that creativity emerges from interactional dynamics rather than isolated acts. The \ac{hci} community has applied these principles to collaborative systems; for example, \citet{kim2014ensemble} developed the Ensemble system, which asymmetrically distributes creative responsibility between leaders and crowds.

In Ensemble~\citep{kim2014ensemble}, lead authors maintain a high-level vision through ``scene prompts'' while contributors generate content within specified boundaries. This suggests that strategic constraints can focus attention on specific narrative elements to support creative freedom. Our system extends this principle by using story beats as navigational constraints and replacing human crowds with specialized \ac{ai} personas. This configuration shifts the leader-contributor dynamic into a human-AI partnership mediated by specific creative voices, allowing for improvisational emergence within a structured framework.

Recent research has examined \ac{ai} capabilities in distributed creative processes. \citet{wang2024unleashing} investigated ``cognitive synergy'' through Solo Performance Prompting (SPP), where a single \ac{llm} adopts multiple collaborative personas. Their results indicated that using specialized personas reduced hallucinations and maintained reasoning capabilities during complex tasks. While SPP was evaluated on task-solving rather than narrative generation, it provides a computational basis for applying distributed creativity to collaborative improvisational storytelling.

\section{Formative Study}

To understand the challenges in human-AI collaborative storytelling, we conducted a formative study involving in-depth interviews with writers of varying expertise. Our objective was to identify specific areas where writers require support, thereby informing the design of an AI-based system for collaborative narrative creation.

\subsection{Methods}

\subsubsection{Participants}
Writers are the primary practitioners of narrative storytelling across different media. We recruited five participants through professional networks to ensure a range of expertise, from emerging writers (2 years) to experienced professionals (15+ years; see \cref{tab:participants}). All participants had used \ac{ai} writing tools previously, primarily for brainstorming, research, and story development.

\begin{table*}[t]
  \centering
  \captionsetup{width=0.9\textwidth}
  \caption{\textbf{Participant demographics and professional background.}This table details the gender, age, occupation, education, and years of writing experience for the five study participants.}
  \label{tab:participants}
  \Description{Table showing demographic information of five participants, including gender, age, occupation, education, and years of writing experience.}
  \small
  \begin{tabularx}{0.9\textwidth}{@{} c c c l l c @{}}
    \toprule
    \textbf{ID} & \textbf{Gender} & \textbf{Age} & \textbf{Occupation} & \textbf{Education} & \textbf{Writing Experience (Years)} \\
    \midrule
    W1 & Male & 34 & Screenwriter & Master of Fine Arts in Film Production & 8 \\
    W2 & Male & 34 & Writer & Master of Arts in Creative Writing & 15+ \\
    W3 & Female & 38 & Film Producer/Screenwriter & Bachelor of Arts in Literature & 10+ \\
    W4 & Female & 27 & Freelance Writer & Master of Arts in Anthropology & 2 \\
    W5 & Male & 32 & Content Creator/Educator & Bachelor of Medicine & 7 \\
    \bottomrule
  \end{tabularx}
\end{table*}

\subsubsection{Procedure}
We conducted semi-structured remote interviews via video conferencing, each lasting approximately one hour. The protocol examined participants' experiences across story development phases: ideation, planning, drafting, and revision. We investigated current \ac{ai} tool usage, specific strategies, challenges, and perspectives on how \ac{ai} could support creative storytelling. Participants discussed collaborative writing experiences and their expectations for maintaining agency in human-AI partnerships.

\subsection{Key Findings}

Analysis revealed four areas where writers seek support in collaborative storytelling:

\textbf{Managing Narrative Structure through Segmented Units.} Participants consistently struggled with managing coherence and creative momentum across extended narratives, particularly when balancing spontaneous creativity with structural organization. W1 described the creative process as involving \textit{``spiral progression,''} where writers \textit{``repeatedly break the simple linear, top-down structure,''}. W4 emphasized that compelling writing contains \textit{``randomness''} that \textit{``cannot be explained by high-probability experiences,''} suggesting that effective structural support must accommodate unexpected creative developments. Writers naturally adopted segmented approaches to manage this complexity. W2 utilized \textit{``story beats''} as discrete structural units for maintaining narrative progression without constraining exploration. W1 noted that while \textit{``short stories don't require such strong structural demands,''} longer works benefit from \textit{``breaking it down into familiar story structures like three-act or eight-sequence formats''} that provide navigational waypoints without dictating creative content.

\textbf{Seeking Diverse Perspectives Beyond Single-Voice Generation.} All participants reported creative limitations with single-voice \ac{ai} systems, perceiving them as producing repetitive content lacking genuinely novel narrative elements. W2 explained that existing \ac{ai} systems \textit{``just continue what you're doing''} rather than bringing \textit{``new beats or new elements''} to the story. Writers valued exposure to multiple perspectives during the creative process. W5 advocated for \textit{``multiple viewpoints''} from different disciplinary backgrounds, while W3 emphasized that individual creative capacity is inherently \textit{``limited''} and benefits from diverse input. W1 described AI's potential strength in \textit{``finding a partner''} that could provide \textit{``different possibilities''} and perform creative \textit{``combinations,''} suggesting effective \ac{ai} storytelling systems should provide diverse creative alternatives.

\textbf{Retaining Creative Ownership during \ac{ai} Assistance.} While appreciating \ac{ai} assistance, participants strongly emphasized maintaining creative ownership and preserving improvisational creativity. W3 articulated this need clearly: \textit{``I am the one making judgments, I am the one making the final decisions... so this is my story.''} However, participants criticized inefficient interaction patterns with current \ac{ai} tools. W2 described the frustration with \textit{``continually prompting it for like half an hour''} without productive outcomes, highlighting the need for more effective collaborative workflows that preserve creative agency while providing needed support. 

\textbf{Coordinating Consistency across Multi-Voice Contributions.} A challenge emerged when managing coherence across multiple creative contributors. W2 highlighted collaboration difficulties, noting how different writing styles can result in \textit{``clashing with each other.''} When multiple voices contribute to a single narrative, maintaining consistent character development, plot logic, and thematic coherence becomes increasingly complex and requires careful coordination beyond individual writing processes. W3 observed AI's limitations in maintaining consistency across extended collaborative works, describing current systems as having limited capacity for coherent long-form generation. This challenge encompasses more than basic fact-checking---it involves coordinating deeper narrative elements including character development arcs, thematic consistency, and tone maintenance throughout extended collaborative creation processes.

\subsection{Design Goals}

Based on these findings, we identified four design goals for human-AI collaborative storytelling:

\textbf{DG1: Structuring Creative Development through Narrative Units.} The system should decompose story creation into discrete, well-defined narrative units (e.g., story beats). These serve as creative waypoints, allowing writers to balance improvisation with structural organization in extended narratives.

\textbf{DG2: Expanding Creative Exploration through Diverse Narrative Voices.} To overcome the repetitive nature of single-model generation, the system should employ multiple specialized creative voices. These voices should provide unexpected directions and diverse perspectives to enhance exploration.

\textbf{DG3: Empowering User Agency through Selective Control.} The system should position users as creative directors who evaluate, select, and refine AI-generated elements. Mechanisms should enhance human decision-making rather than automate the creative process entirely.

\textbf{DG4: Supporting Narrative Coherence across Collaborative Inputs.} The system should monitor and support narrative coherence across diverse contributions. It must help writers identify and resolve inconsistencies in character arcs and thematic elements to ensure quality throughout the collaborative process.

\section{System Design} \label{sec:system}

\subsection{Theoretical Foundation}

Based on our formative study, we identified the need for a framework that balances creative diversity with human agency and narrative coherence. We adopted Campbell’s \acf{bvsr} model~\citep{campbell1960blind}, which applies evolutionary principles to creativity. \ac{bvsr} posits that creative processes require two distinct phases: the generation of diverse alternatives (blind variation) and the systematic evaluation and retention of promising options (selective retention).

\textbf{Blind Variation}: This phase involves generating alternatives independently of existing patterns or statistical likelihoods. This prevents the system from converging on predictable outputs, which is a known limitation of next-token prediction in \acp{llm}. By implementing variation through specialized personas, the system generates diversity at structural and causal levels rather than only varying surface-level linguistics. This approach directly supports \textit{Design Goal 2}. 

\textbf{Selective Retention}: In this phase, promising variations are evaluated and retained. Because this requires contextual understanding and domain expertise, the system assigns this role to the user. This maintains human agency by positioning the user as the primary decision-maker, supporting \textit{Design Goal 3}. Within this framework, the \ac{ai} provides generative support while the human user guides the narrative trajectory.

\textbf{Iterative Cycles}: \ac{bvsr} suggests that creativity emerges through repeated variation-selection cycles. Each selection provides the narrative context for subsequent variations. This informs our beat-based architecture \textit{(Design Goal 1)}, where the system generates multiple variations for each story beat, and human selection guides narrative progression. This structure supports \textit{Design Goal 4} by enabling coherence through the interplay between algorithmic variation and human curation, rather than relying solely on predefined narrative constraints.

\subsection{Design Principles}

This section details the design decisions used to implement \ac{bvsr} theory within a co-creative storytelling system.

\subsubsection{User Workflow Design}

\begin{figure*}[htbp]
    \centering
    \includegraphics[width=\linewidth]{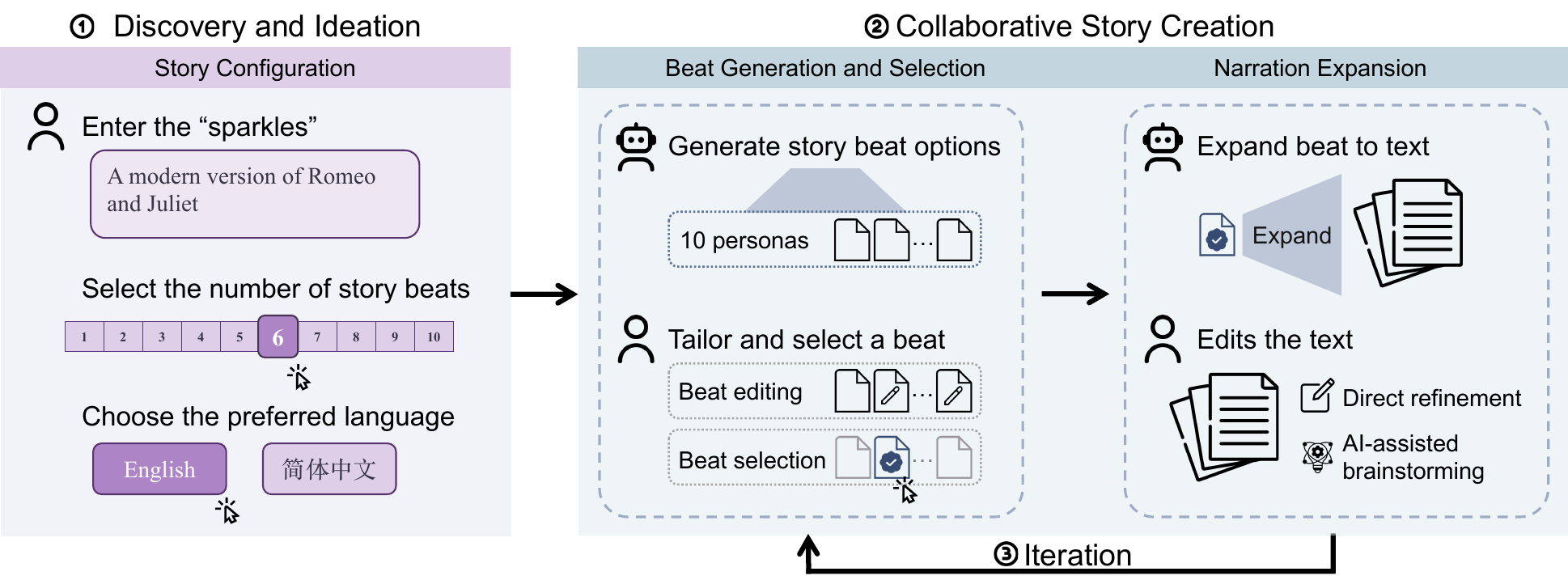}
    \caption{\textbf{The user workflow of \model.} The process consists of three integrated phases: (i) \textbf{Discovery and Ideation}, where users initialize the narrative by entering ``sparkles'' and selecting parameters such as language and story length; (ii) \textbf{Collaborative Story Creation}, where the system generates diverse beat options using 10 distinct personas, allowing users to tailor, select, and expand beats into full narrative, and edit the text  before iterating to the next story beat; and (iii) \textbf{Iteration}, where users build their story progressively by repeating the beat selection and narrative expansion process.}
    \Description{A workflow diagram with three main phases shown as connected boxes. Phase one shows an input interface where users enter story ideas and parameters. Phase two displays a branching structure where ten different \ac{ai} personas generate alternative story beats, with the user selecting one option from the alternatives. The selected beat expands into a full narrative text with editing capabilities. Phase three shows this process repeating cyclically, with each iteration building upon previous story segments to create a progressively longer narrative.}
    \label{fig:workflow} 
\end{figure*}

\model uses a three-phase workflow (see \cref{fig:workflow}) based on \textbf{story beats}---discrete units of narrative progression containing settings, characters, and events. In screenwriting, beats function as the fundamental components of narrative arcs~\citep{newman2006beats}, making them an effective unit for variation-selection dynamics.

\textbf{Discovery and Ideation:} Users provide initial narrative inputs (``sparkles'') and define story parameters. This establishes the initial state while preserving a broad exploration space for subsequent variation.

\textbf{Collaborative Story Creation:} This phase implements variation through multi-persona generation. Ten specialized personas generate beat alternatives simultaneously. The interface displays the structural details and rationale for each proposal. Users then perform selective retention by evaluating these alternatives against their narrative goals. Users can modify selected beats before expanding them into 800--1000 word prose.

\textbf{Iteration:} Users repeat the generation and selection process to build the narrative. The system integrates previous context while maintaining generative diversity. A \acs{rag}-based consistency system encodes story history into semantic embeddings. When new beats are generated, this mechanism identifies logical 
inconsistencies and adjusts rankings to prioritize coherent 
options while preserving alternatives.

\subsubsection{Story Beat Architecture [DG1]}

\paragraph{Design Rationale.} 

To maintain manageability, storytelling is structured into discrete \textit{story beats}. Our design prioritizes semantic transparency by decomposing each narrative segment into setting, characters, and key events. This modularity allows for precise modifications---such as changing a specific character action---without requiring the regeneration of the entire scene. This granularity provides clear causal anchors for the model and ensures contextual coherence.

\subsubsection{Multi-Persona Generation for Creative Diversity [DG2]}

\paragraph{Design Rationale.}

To increase output diversity, the system uses genre-based rather than style-based personas. Style-oriented approaches primarily affect lexical choice, whereas genre-based personas influence narrative logic and causal structures. Genre dictates the types of events and the underlying logic of the narrative. For example, a \textit{Mystery} persona integrates structural elements such as information asymmetry and the strategic placement of clues.

\paragraph{Persona Design.}

We selected ten personas based on three criteria:
\begin{itemize}[leftmargin=*]
    \item \textbf{Genre Coverage}: Representation across major narrative categories grounded in established literary frameworks~\citep{earnshaw2014handbook}.
    \item \textbf{Narrative Differentiation}: Personas span different approaches, including plot-driven (\eg, \textit{Adventure Guide}), character-centric (\textit{Romance Matchmaker}), and world-building roles (\textit{Fantasy World Builder}).
    \item \textbf{Complementary Functions}: Personas emphasize different narrative elements, such as atmosphere (\textit{Horror Atmosphere Creator}) or social commentary (\textit{Dystopian Visionary}).
\end{itemize}

This diversity enables \model to explore different regions of the narrative space simultaneously, rather than clustering around a single narrative trajectory with minor variations. The complete persona specifications are detailed in \cref{tab:personas}.

\begin{table*}[htbp]
    \centering
    \setlength{\tabcolsep}{4pt}
    \small
    \captionsetup{width=0.9\textwidth}
    \caption{\textbf{Storyteller personas in \model.} Ten specialized storytelling personas form the generative ensemble of \model, each designed with distinct genre expertise and narrative capabilities.}
    \label{tab:personas}
    \begin{tabularx}{0.9\textwidth}{lX}
        \toprule
        \textbf{Persona} & \textbf{Narrative Specialization} \\
        \midrule
        Fantasy World Builder & Specialize in crafting rich and imaginative fantasy worlds, complete with intricate magic systems, mythical creatures, and diverse cultures. \\
        Sci-Fi Futurist & Focus on creating believable and innovative science fiction settings, incorporating advanced technology, space travel, and futuristic societies. \\
        Mystery Solver & Assist in developing complex and intriguing mysteries, helping to plant clues, red herrings, and plot twists that keep readers engaged until the narrative resolution. \\
        Romance Matchmaker & Skilled at creating compelling romantic storylines, ensuring that character chemistry feels authentic and that relationships develop naturally over narrative progression. \\
        Historical Researcher & Excel at incorporating accurate historical details and context into narratives, bringing historical fiction to life and immersing readers in specific temporal settings. \\
        Horror Atmosphere Creator & Help to build tension and suspense in horror narratives, using descriptive language and pacing to create unsettling atmospheric elements that enhance reader engagement. \\
        Adventure Guide & Specialize in crafting thrilling adventure stories, designing exciting action sequences, perilous obstacles, and high-stakes challenges for character development. \\
        Comedy Humorist & Focus on incorporating humor and wit into narratives, using wordplay, situational comedy, and character interactions to enhance narrative enjoyment. \\
        Dystopian Visionary & Adept at constructing dystopian settings and exploring societal and political implications, helping to create thought-provoking and cautionary narrative frameworks. \\
        Magical Realism Conjuror & Assist in blending fantastical elements with everyday reality, creating narratives that are simultaneously grounded and whimsical, featuring extraordinary occurrences within otherwise ordinary contexts. \\
        \bottomrule
    \end{tabularx}
\end{table*}

\subsubsection{Selective Retention Interface for Creative Agency [DG3]}

\paragraph{Design Rationale.}

The interface positions users as the final evaluators through a multi-layered design. This supports different levels of engagement based on the user's specific creative goals.

\paragraph{Interaction Design.}

\model preserves agency through three patterns:
\begin{itemize}[leftmargin=*]
    \item \textbf{Direct Selection:} This design supports a fluid drafting flow by allowing writers to adopt preferred beats with a single click. This mode is designed for moments of high creative momentum, where the priority is rapid progression rather than granular deliberation.
    \item \textbf{Comparative Evaluation:} This design encourages reflective decision-making through side-by-side comparison. To minimize cognitive load, the interface presents two alternatives simultaneously while providing access to full candidates through progressive disclosure.
    \item \textbf{Multi-stage Editing:} This design treats AI-generated suggestions as provocative starting points rather than finalized text. This design allows writers to refine both individual story beats and expanded narratives, ensuring that the final output remains grounded in the author's unique voice and intent.
\end{itemize}

\subsubsection{Consistency Management Through Soft Constraints [DG4]}

\paragraph{Design Rationale.}

Managing coherence requires balancing narrative stability with creative variation. Rather than using hard constraints that automatically filter out inconsistent options, we implemented a ranking system. This prioritizes logically consistent paths while allowing users to choose divergent options if they serve a specific creative purpose.

\subsection{Technical Implementation}

\subsubsection{Beat Generation Pipeline}

\paragraph{Technical Architecture.} 

\begin{figure*}[htbp]
    \centering
    \includegraphics[width=\linewidth]{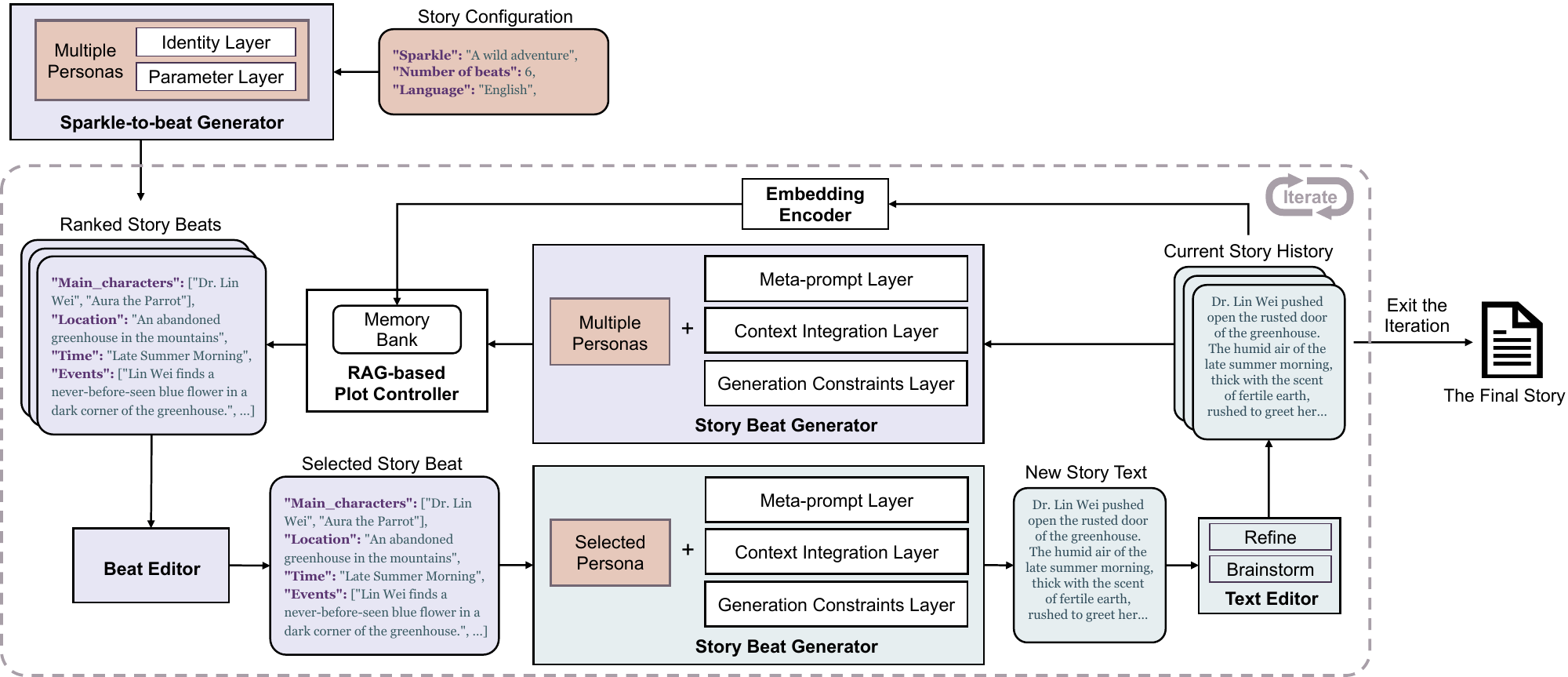}
    \caption{\textbf{\model's technical pipeline.} The system transforms user sparkles into story beats via multi-persona generation, employing a three-layer prompt architecture (meta-prompt, context integration, generation constraints) across both beat and text generation stages. The \acs{rag}-based Plot Controller ensures narrative consistency while users iterate through beat selection and text refinement. Purple indicates beat-level operations and green indicates text-level operations. Rectangles represent functional modules, while rounded corners represent data.}
    \Description{A technical pipeline flowchart showing data flow from left to right. User input sparkles enter a multi-persona generation module (shown in purple), where three prompt layers process the input. Ten parallel processing streams generate diverse story beats. A \acs{rag}-based plot controller module checks consistency and ranks options. The user selects a beat, which feeds into a text generation module (shown in green) that expands the beat into narrative prose. Arrows indicate iterative loops where generated text feeds back into the system for subsequent beat generation. Rectangular boxes represent processing modules, while rounded rectangles represent data states.}
    \label{fig:pipeline} 
\end{figure*}

The beat generation employs a three-layer prompt architecture (see \cref{fig:pipeline}). The \textbf{meta-prompt layer} establishes the collaborative storytelling framework for the multi-persona system. The \textbf{context integration layer} combines compressed story history with the current beat state to maintain narrative continuity. The \textbf{generation constraint layer} specifies structural requirements and coherence criteria for outputs.

\paragraph{Data Schema and Structured Output.} 

Each story beat is generated as a structured JSON object with three 
key fields: \texttt{setting} for spatio-temporal context, \texttt{characters} for active participants, and \texttt{key\_events} for 3--5 pivotal actions. This structured format enables consistent processing while remaining interpretable during user selection. The system defaults to a six-beat structure following \citet{hauge2011writing}'s framework, though users can configure 1--10 beats. Beat complexity adapts to narrative position: initial beats contain 3--4 events for world-building, while climactic beats expand to 4--5 events for dramatic intensity. Users can further adjust complexity through interactive refinement.

\paragraph{Stage-Specific Generation Logic.}

Sparkle-to-beat generation transforms the user's initial narrative seed into the first story beat, while subsequent beat generation builds upon established story elements. 

\paragraph{Text Expansion Process.}

When the user selects a beat, the system expands it into 800--1000 words of narrative text. The expansion uses the same three-layer prompt architecture, with the selected persona passed as a parameter. This parameterized approach avoids the need for multiple persona-specific generators, reducing both cost and latency. Throughout generations, the system maintains up to 8000 tokens of narrative history to preserve context continuity.

\subsubsection{Persona Instantiation and Parallel Generation}

\paragraph{Prompt-Based Persona Instantiation.}

Each persona is defined through a multi-layered prompt template. The \textbf{identity layer} establishes genre-specific writing philosophies and creative priorities, while the \textbf{parameter layer} defines quantitative constraints on narrative elements (e.g., lexical diversity, dialogue-to-narrative ratios). Rather than providing static content instructions, this structure shapes the persona's generative behavior at a deeper level. Each persona is instantiated as a separate call to GPT-4o, with these layers driving consistent persona-specific variation across generations.

\paragraph{Parallel API Coordination.}

Rather than sequential generation, all ten personas generate story beats simultaneously through parallel API calls. This parallel architecture ensures that each persona produces independent alternatives without influence from other outputs. Each API call includes the complete story context (up to 8000 tokens of narrative history) and persona-specific instructions (approximately 500 tokens of role-defining prompts).

\subsubsection{Interface Architecture and Interaction Mechanisms}

\begin{itemize}[leftmargin=*]
    \item \textbf{Brainstorm Module:} This module facilitates open-ended exploration by allowing users to interact with the \ac{ai} to generate narrative inspiration. It serves as a conversational partner to help users overcome creative blocks, explore alternative plot developments, or elaborate on character motivations without directly modifying the current draft.
    \item \textbf{Refine Module:} This module provides direct editorial support by allowing users to submit specific revision requests. Users can instruct the \ac{ai} to rewrite existing prose to adjust pacing, enhance descriptive detail, or modify dialogue for better character consistency. This ensures that the final text reflects the user's specific stylistic and narrative intent through iterative revision.
\end{itemize}

\begin{figure*}[htbp]
    \centering
    \begin{subfigure}[t]{0.475\linewidth}
        \includegraphics[width=\linewidth]{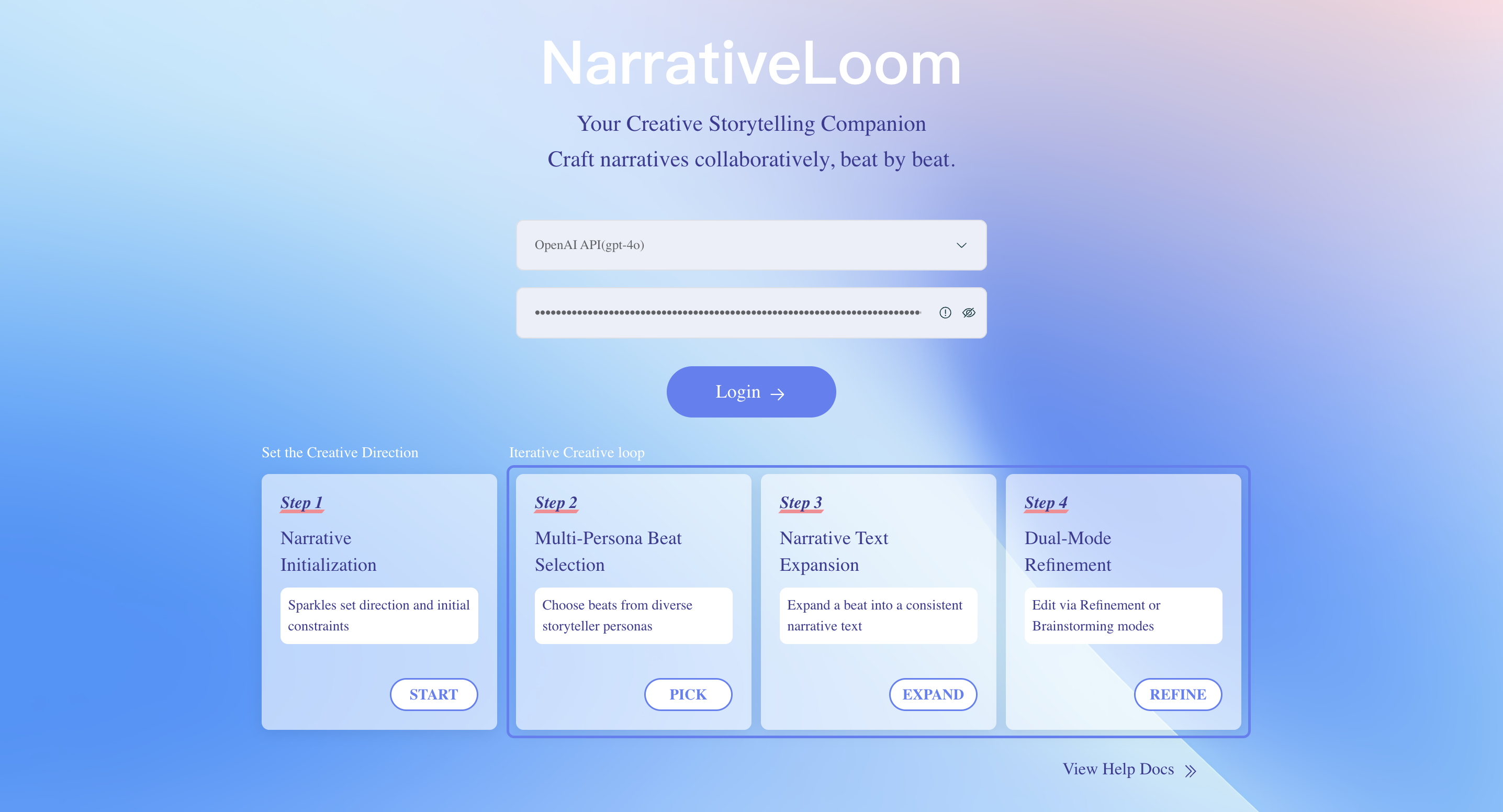}
        \caption{Narrative initialization interface}
    \end{subfigure}%
    \begin{subfigure}[t]{0.475\linewidth}
        \includegraphics[width=\linewidth]{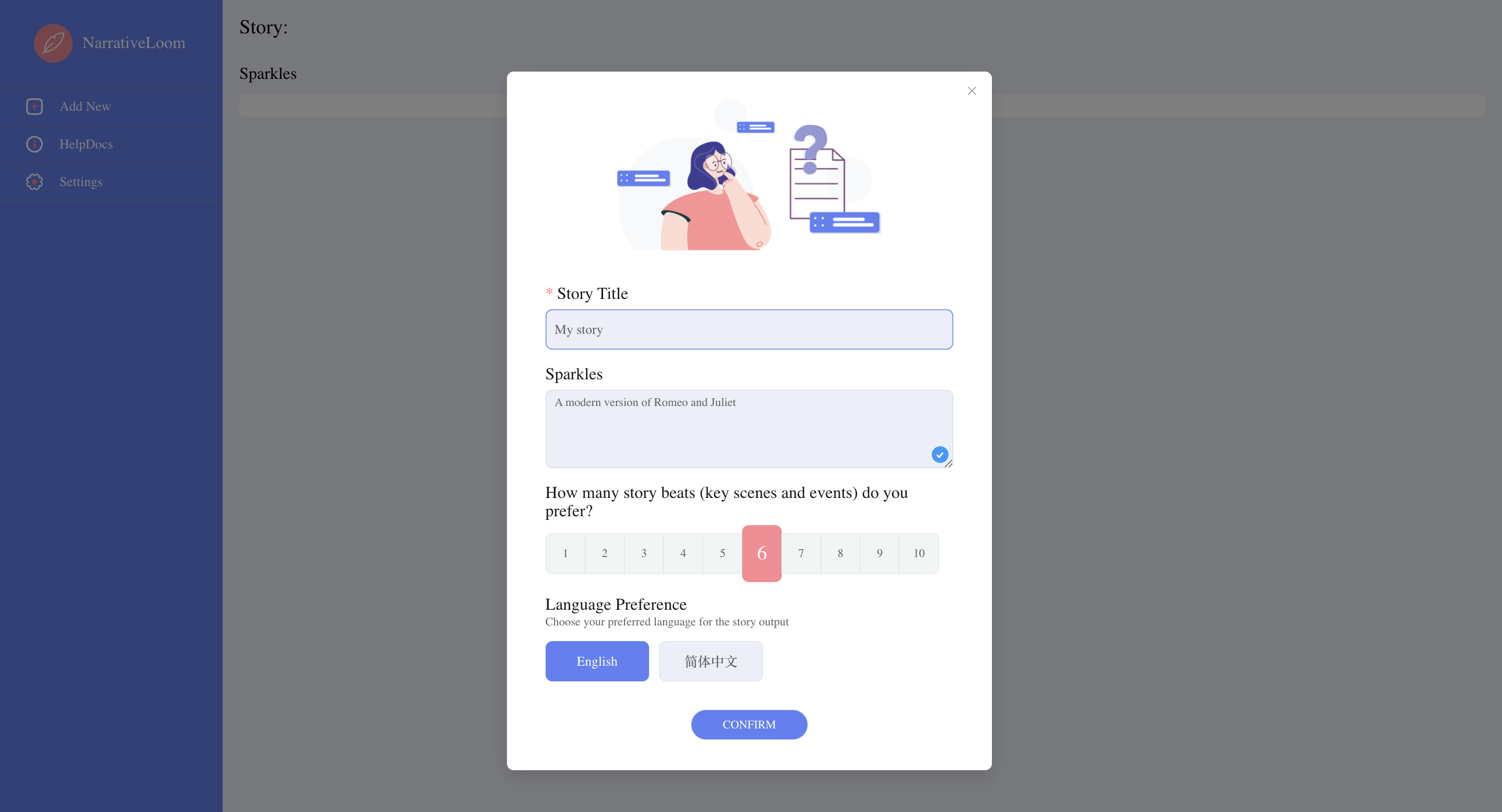}
        \caption{Generation parameter configuration}
    \end{subfigure}%
    \\%
    \begin{subfigure}[t]{0.475\linewidth}
        \includegraphics[width=\linewidth]{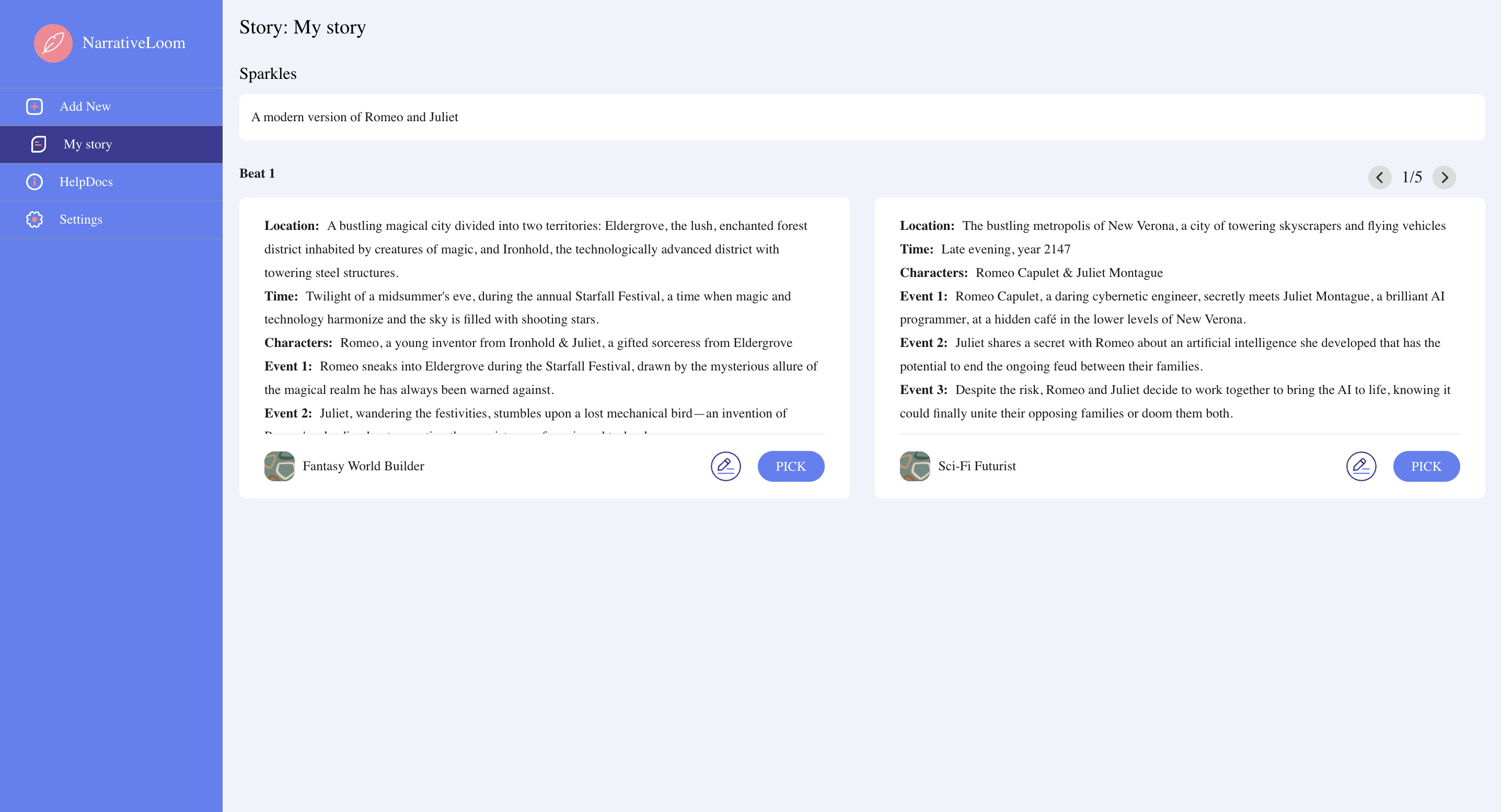}
        \caption{Multi-persona beat selection}
    \end{subfigure}%
    \begin{subfigure}[t]{0.475\linewidth}
        \includegraphics[width=\linewidth]{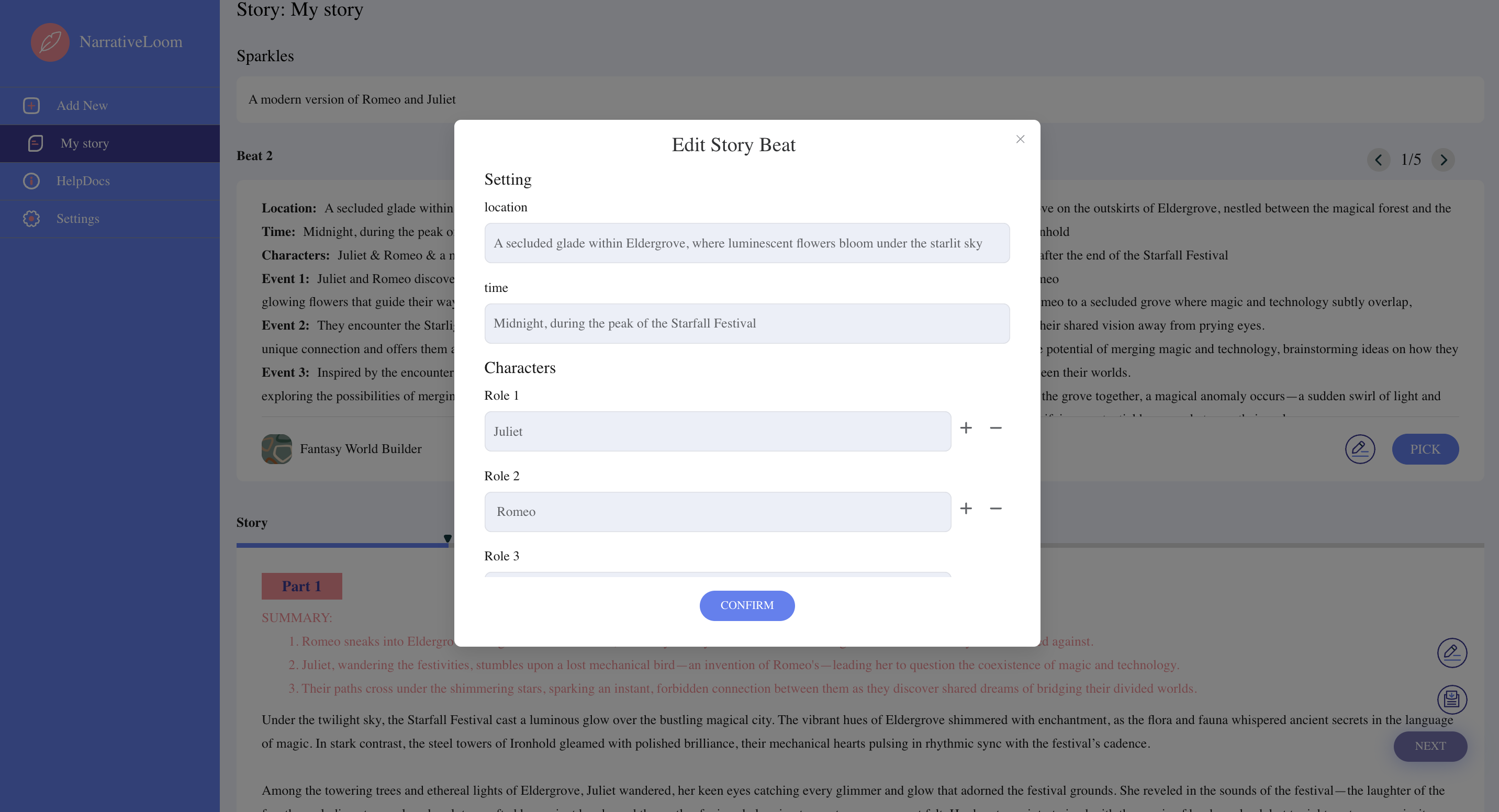}
        \caption{Structural beat modification interface}
    \end{subfigure}%
    \\%
    \begin{subfigure}[t]{0.475\linewidth}
        \includegraphics[width=\linewidth]{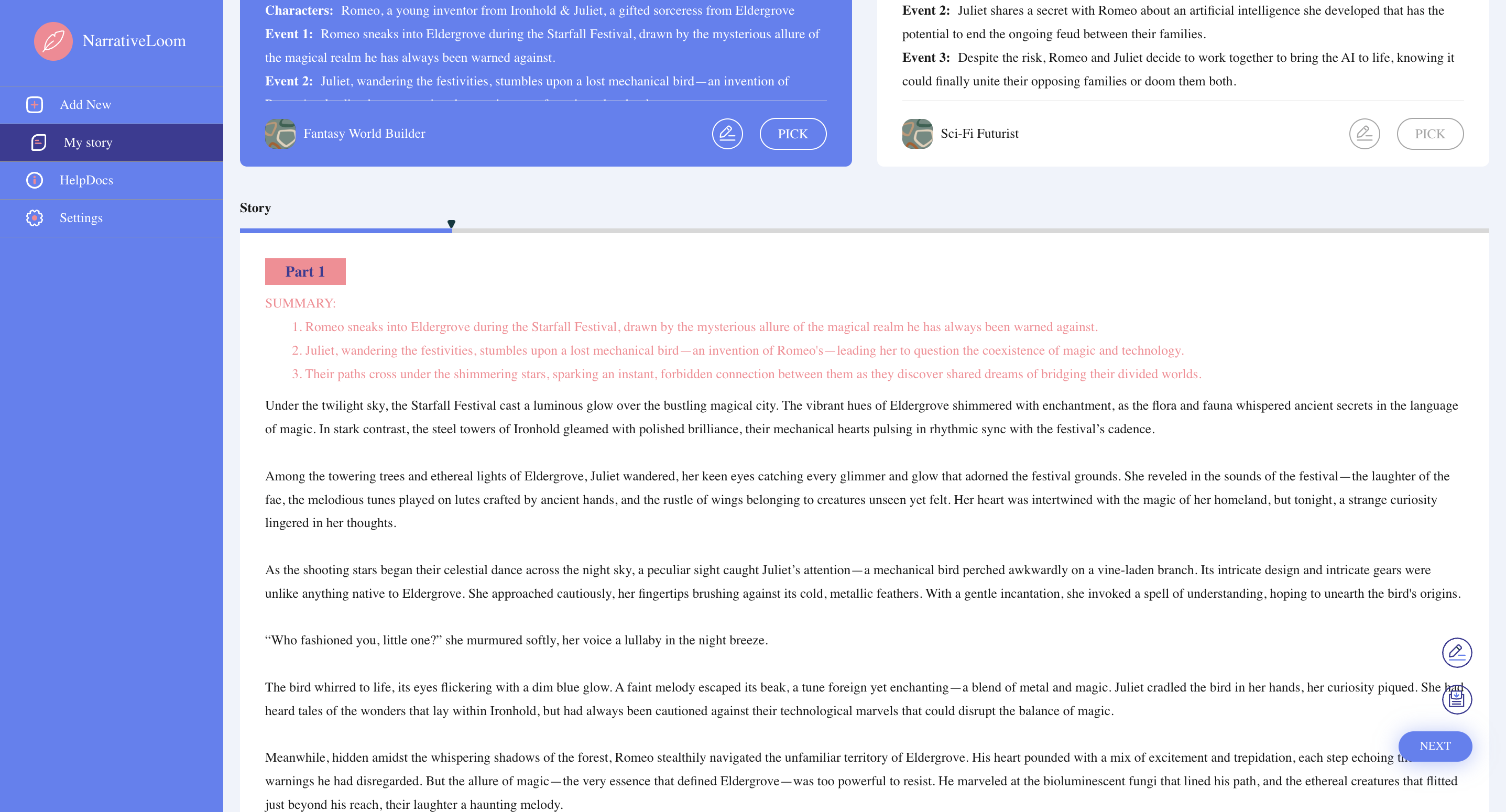}
        \caption{Narrative text expansion}
    \end{subfigure}%
    \begin{subfigure}[t]{0.475\linewidth}
        \includegraphics[width=\linewidth]{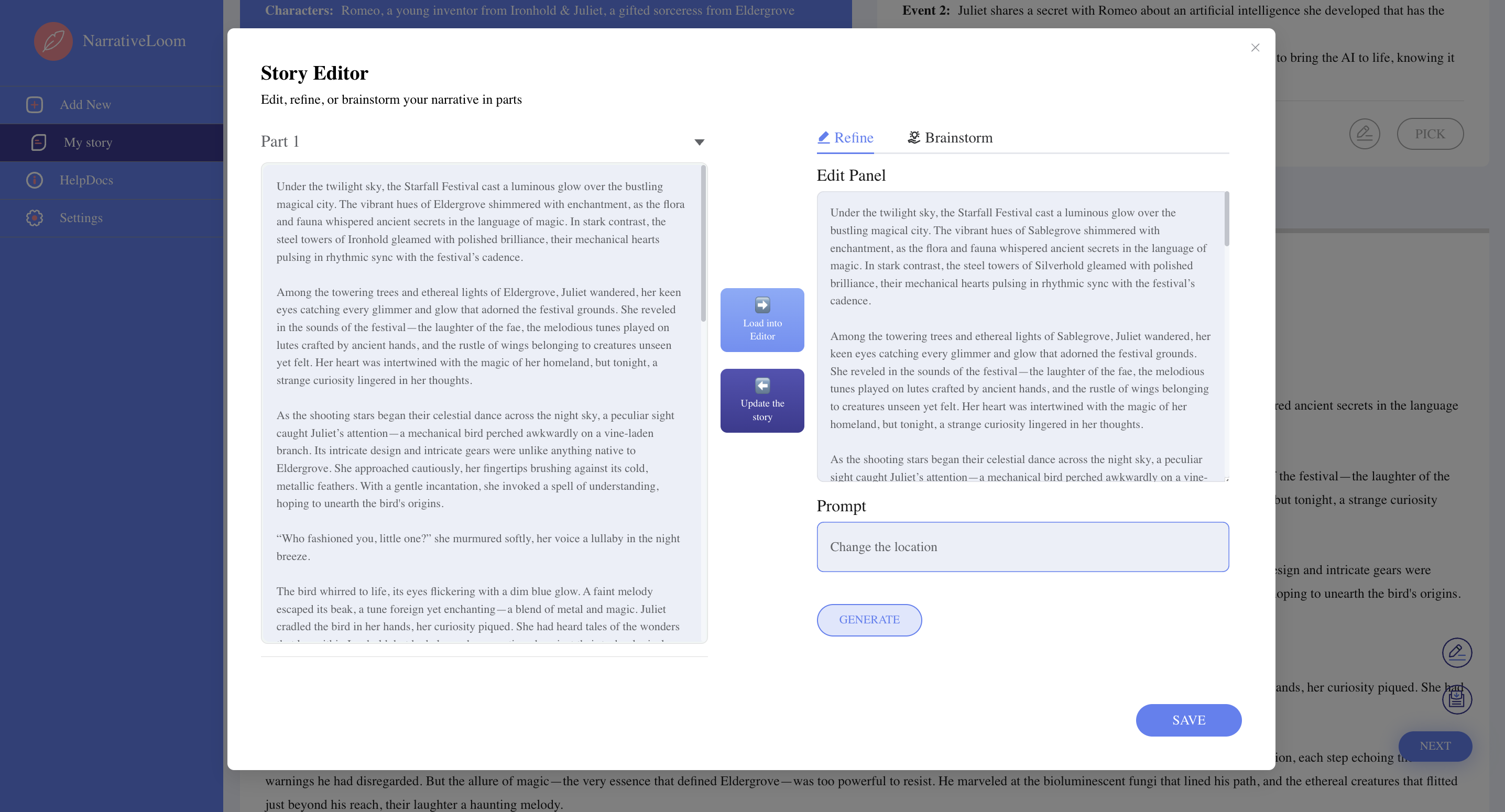}
        \caption{Dual-mode refinement mechanism}
    \end{subfigure}%
    \caption{\textbf{The \model interface implementing the \acs{bvsr} framework.} The six panels illustrate the system's functional components: (a) narrative initialization interface; (b) generation parameter configuration; (c) multi-persona beat selection; (d) structural beat modification interface; (e) narrative text expansion; and (f) dual-mode refinement mechanism. This workflow supports iterative variation and selection while preserving human creative agency.}
    \Description{A collection of six screenshots demonstrating the \model system interface. Panel (a) shows a text input field for narrative seeds. Panel (b) displays sliders and dropdown menus for selecting story length and language. Panel (c) presents a grid of ten persona-based beat options with associated rationales. Panel (d) shows an editable table where users can modify settings, characters, and events of a selected beat. Panel (e) displays a large text area where the beat is expanded into full narrative prose. Panel (f) features a sidebar with two primary tools: a brainstorm module for conversational ideation and a refinement module for direct text revision.}
    \label{fig:interface-workflow}
\end{figure*}

\paragraph{Interface Overview.}

The interface is built using Streamlit and employs a three-panel layout for multi-persona narrative exploration (see \cref{fig:interface-workflow}). The left sidebar manages story portfolios, the central workspace displays persona-generated content for comparison, and the lower panel supports narrative refinement through dual-mode editing. In the central workspace, each persona's beat proposal appears in a dedicated panel showing its role-specific rationale and templated content. The system maintains active session data in memory while persisting user story portfolios to storage.

\paragraph{Beat Selection Display.}

The central workspace ranks beat proposals by consistency using the RAG-based plot controller (detailed in \cref{sec:rag-implementation}). The interface displays consistent options first and keeps all ten persona outputs accessible for exploration.

\paragraph{Editing and Refinement.}

The system supports editing at two levels to ensure human-AI alignment throughout the creative process. \textbf{Beat-level editing} allows users to modify settings, characters, and events before text expansion, with these structural changes propagating through the generation pipeline. \textbf{Text-level editing} enables prose refinement through manual intervention or two specialized AI-powered modules:

\subsubsection{\acs{rag}-Based Plot Controller Implementation}\label{sec:rag-implementation}

\paragraph{Technical Implementation.}

The plot controller maintains narrative coherence through a consistency check system utilizing \acf{rag} via LlamaIndex~\citep{liu2022llamaindex}. The system detects logical inconsistencies between newly generated beats and established story elements by combining semantic retrieval with language model reasoning.

\paragraph{Story History Indexing.}

Story history is indexed via a vector-based memory mechanism. Each completed story segment is treated as a standalone document and encoded using the OpenAI \texttt{text-embedding-ada-002} model. This process generates 1536-dimensional embeddings that represent the semantic content of previous narrative developments. These embeddings are stored in a LlamaIndex vector store, allowing the system to retrieve relevant historical context during the evaluation of subsequent story beats.

\paragraph{Consistency Check Process.}

For each new story beat generated by a persona, the system executes contextual retrieval and logical verification. The beat content is converted into a query, embedded, and compared against the vector store using cosine similarity:
\[
    \text{sim}(\mathbf{v}_{\text{beat}}, \mathbf{v}_i) = \frac{\mathbf{v}_{\text{beat}} \cdot \mathbf{v}_i}{\|\mathbf{v}_{\text{beat}}\| \|\mathbf{v}_i\|},
\]
where $\mathbf{v}_{\text{beat}}$ is the embedding of the verification query and $\mathbf{v}_i$ represents the $i$-th story history embedding. The retrieved contexts are provided to GPT-3.5-turbo with the following prompt: \textit{``The new story beat is: [beat content]. Are there any logical errors in the events of the new story beat? Answer briefly in [Yes] or [No]. If [Yes], briefly describe the errors.''} The \ac{llm} evaluates the beat for contradictions regarding prior events, character status (\eg, unexplained revivals), temporal sequences, or established world rules. If an error is identified, the system does not discard the beat but labels the inconsistency for the user. 

\section{User Study}

We conducted a within-subjects user study to evaluate the effectiveness of \model in supporting collaborative storytelling. Our evaluation compared the proposed system against a conventional single-persona chatbot interface to investigate its influence on the creative process, narrative quality, and user engagement. We hypothesized that the structured guidance and diverse narrative options provided by \model would enhance storytelling capabilities relative to the baseline chatbot.

\subsection{Participants}

Participants were recruited via Prolific, with eligibility restricted to native English speakers over 18 years of age holding at least an undergraduate degree. This criterion was established to ensure sufficient proficiency for complex narrative tasks. Of the 109 initial recruits, 94 passed a mandatory familiarization test, which served as a procedural fact-check to verify comprehension of the interface mechanics and core narrative rules. 

To maintain high data quality, participants were excluded if they spent less than 20 minutes interacting with the systems or completed fewer than two story beats in the \model condition. This filtering process resulted in a final sample of \textbf{50 participants} for analysis. The final cohort (24 female, 26 male) had a mean age of 34.82 years ($SD = 10.03$, range: 22--71). Reported ethnicities included White (n=35), Asian (n=7), Black (n=6), Mixed (n=1), and Non-disclosure (n=1). 

Diverse educational backgrounds were represented, including Arts and Humanities (n=11), Business (n=11), Natural Science (n=10), Social Science (n=10), and other fields (n=8). Regarding degree attainment, 33 participants held undergraduate degrees, 15 held graduate degrees, and 2 held doctoral degrees. Compensation was provided at a rate of £9/hour. The study protocol was approved by the university's Institutional Review Board (IRB).

\subsection{Study Procedure and Protocol}

We employed a within-subjects design in which participants interacted with both the experimental system (\model) and a baseline system. To mitigate order effects, system presentation was counterbalanced across participants~\citep{perreault1975controlling}. The study followed five sequential phases:

\begin{enumerate}[leftmargin=*,label=\arabic*.]
    \item \textbf{Demographics and Background (10 min):} Participants reported demographic data, writing experience, and genre preferences. 
    \item \textbf{Creative Warm-up (5 min):} Participants drafted initial story ideas (``sparkles'') and wrote a short story for two minutes to establish a baseline for unaided creation.
    \item \textbf{Familiarization and Training (10 min):} Participants reviewed an infographic of the protocol and completed two comprehension questions to verify task understanding.
    \item \textbf{System Interaction (40 min):} Participants used each system for 20 minutes following a 3-minute tutorial per system.
    \begin{itemize}[leftmargin=*]
        \item \textbf{\model\ Condition (20 min):} Participants used our multi-persona system (GPT-4o API). For each narrative round, the system generated multiple continuations via anonymized \ac{ai} personas. Users selected, edited, or built upon these options without knowing the specific persona source. Participants were required to complete at least two story beats using their sparkles.
        \item \textbf{Chatbot Condition (20 min):} As a control, participants interacted with a custom chatbot interface powered by the same model using the same sparkle. Stories were developed through direct conversational prompting with a single \ac{ai} agent.
    \end{itemize}
    \item \textbf{Post-Task Evaluation (10 min):} After each condition, participants completed a survey regarding their experience with the systems.
\end{enumerate}

\subsection{Data Collection and Analysis}

\begin{table*}[htbp]
  \centering
  \captionsetup{width=0.9\textwidth}
  \caption{\textbf{Expert evaluator demographics and professional experience.} Profiles of the four expert reviewers, including their occupation, education, and years of professional writing experience.}
  \label{tab:expert_participants}
  \Description{Demographic information of four expert evaluators, including their creative roles, advanced degrees in film and literature, and extensive writing experience ranging from 8 to 18 years.}
  \small
  \begin{tabularx}{0.9\textwidth}{@{} c c c l l c @{}}
    \toprule
    \textbf{ID} & \textbf{Gender} & \textbf{Age} & \textbf{Occupation} & \textbf{Education} & \textbf{Writing Experience (Years)} \\
    \midrule
    E1 & Female & 32 & Screenwriter & Master of Fine Arts in Film Production & 12 \\
    E2 & Male & 34 & Screenwriter and Director & Master of Fine Arts in Film Production & 8 \\
    E3 & Female & 38 & Creative Producer & Bachelor of Arts in Chinese Literature & 18 \\
    E4 & Male & 31 & Journalist and Writer & Bachelor of Arts in Comparative Literature & 10 \\
    \bottomrule
  \end{tabularx}
\end{table*}

\begin{table*}[htbp]
    \centering
    \captionsetup{width=0.9\textwidth}
    \caption{\textbf{\ac{ttcw} evaluation framework.} A systematic assessment consisting of 14 binary criteria across four creative dimensions (Fluency, Flexibility, Originality, and Elaboration) to evaluate narrative quality.}
    \label{tab:ttcw-tests}
    \small
    \renewcommand{\arraystretch}{1.1}
    \begin{tabularx}{0.9\linewidth}{@{}lp{10cm}@{}}
    \toprule
    \textbf{Dimension} & \textbf{Test Questions} \\
    \midrule
    \multirow{5}{*}{\textbf{Fluency}} 
    & Does the manipulation of time feel appropriate and balanced? \\
    & Does the story display awareness of balance between scene and summary? \\
    & Does the story make sophisticated use of metaphor or literary devices? \\
    & Does the end feel natural and earned, rather than arbitrary? \\
    & Do story elements work together to form a unified whole? \\
    \midrule
    \multirow{3}{*}{\textbf{Flexibility}} 
    & Does the story provide diverse, convincing perspectives? \\
    & Does the story balance interiority and exteriority effectively? \\
    & Does the story contain turns that are both surprising and appropriate? \\
    \midrule
    \multirow{3}{*}{\textbf{Originality}} 
    & Will readers obtain unique ideas from this story? \\
    & Is the story original without clichés? \\
    & Does the story show innovation in structure or format? \\
    \midrule
    \multirow{3}{*}{\textbf{Elaboration}} 
    & Does the writer make the fictional world believable at the sensory level? \\
    & Are characters developed at appropriate complexity levels? \\
    & Does the story operate at multiple levels of meaning? \\
    \bottomrule
    \end{tabularx}
\end{table*}

We collected quantitative and qualitative data to evaluate the creative product, creative process, and narrative quality.

\subsubsection{Creative Product and Process}

We measured the quality of the creative product and process using metrics adapted from the Creative Product Semantic Scale (CPSS)~\citep{o1989development} and the Creativity Support Index (CSI)~\citep{carroll2009creativity, cherry2014quantifying}. These adaptations were necessary because the original instruments were designed for different creative domains---CPSS for general creative products and CSI for broader creativity support tools---requiring domain-specific modifications for storytelling evaluation.

Our adaptations involved three key modifications:

\begin{enumerate}[leftmargin=*]
    \item \textbf{Domain Specialization:} We translated abstract creativity dimensions into storytelling-specific constructs. For instance, CPSS's general ``resolution'' dimension was operationalized as narrative ``coherence'' (absence of plot inconsistencies), while CSI's ``exploration'' was reframed as narrative ``diversity'' (range of story possibilities).
    \item \textbf{Scale Simplification:} We streamlined the original 71-item CPSS and 12-item CSI into six focused dimensions using 5-point Likert scales, reducing participant fatigue while maintaining construct validity for comparative evaluation.
    \item \textbf{Contextual Relevance:} We reformulated questions to reflect AI-assisted storytelling contexts. For example, CSI's ``collaboration'' factor was adapted to ``customization,'' measuring the system's ability to align AI-generated content with users' creative vision rather than human-human collaboration.
\end{enumerate}

Participants rated their experience across six dimensions organized into two categories:

\begin{itemize}[leftmargin=*]\label{metric}
    \item \textbf{Product Metrics:} 
    \begin{itemize}
        \item \textit{Novelty} (adapted from CPSS's ``originality''): the degree of surprise and originality in generated narratives
        \item \textit{Diversity} (adapted from CSI's ``exploration''): the range and variety of narrative possibilities offered
        \item \textit{Coherence} (adapted from CPSS's ``resolution''): logical consistency and absence of plot contradictions
    \end{itemize}
    \item \textbf{Process Metrics:}
    \begin{itemize}
        \item \textit{Customization} (adapted from CSI's ``collaboration''): the system's responsiveness to user preferences and creative direction
        \item \textit{Engagement} (adapted from CSI's ``enjoyment''): sustained interest and involvement during story creation
        \item \textit{Usability} (adapted from CSI's ``results worth effort''): ease of interaction relative to creative output quality
    \end{itemize}
\end{itemize}

\subsubsection{Narrative Quality Analysis}

To complement subjective ratings, we conducted computational text analysis of the generated stories using spaCy~\citep{honnibal2020spacy} and NLTK~\citep{bird2009natural}. We analyzed four properties:
\begin{itemize}[leftmargin=*]
    \item \textbf{Word Count:} A basic metric of narrative length and development.
    \item \textbf{Gunning Fog Index:} A measure of text readability and complexity.
    \item \textbf{Dialogue Ratio:} The proportion of text presented as character dialogue \vs narrative exposition.
    \item \textbf{Location Count:} The number of unique spatial references, indicating the richness of the setting.
\end{itemize}

\subsubsection{Persona Selection Tracking.} 

For the \model condition, we tracked which \ac{ai} personas users selected throughout the story creation process. This allowed us to analyze patterns in persona preference, frequency, and selection sequences.

\subsection{Expert Review and Feedback} 
\label{expert_review}

To validate our findings, we recruited four experts with extensive creative writing experience to evaluate the narrative quality of the generated stories (see  \cref{tab:expert_participants} for demographics). The experts worked in two pairs (E1-E2, E3-E4). Each pair independently evaluated the same set of 10 story pairs (20 stories total) randomly sampled from our user study dataset, resulting in a total of 20 pairs and 40 stories evaluated across both expert pairs. To mitigate bias, all stories were anonymized, and their presentation order was randomized.

For the evaluation, we used the \ac{ttcw} protocol~\citep{chakrabarty2024art}, which provides a systematic assessment across four creativity dimensions through 14 binary tests (see \cref{tab:ttcw-tests}). Each expert independently provided a binary pass or fail assessment for each test, accompanied by a brief justification. A story's final creativity score is the total number of tests passed (0--14). Following the \ac{ttcw} evaluation, experts made a forced-choice comparison by selecting what they judged to be the superior story in each pair.

Finally, we conducted semi-structured interviews with all experts to gather qualitative insights on the perceived differences between the systems. The interviews explored five key areas: (i) noticeable differences in creative quality between systems, (ii) \ac{ttcw} dimensions showing the largest quality gaps, (iii) patterns distinguishing higher- \vs lower-quality stories, (iv) recurring creative failures or limitations, and (v) priorities for improving \ac{ai} storytelling systems. We recorded and thematically analyzed the interviews to identify common patterns across expert perspectives.

\begin{figure*}[htbp]
    \centering
    \includegraphics[width=\linewidth]{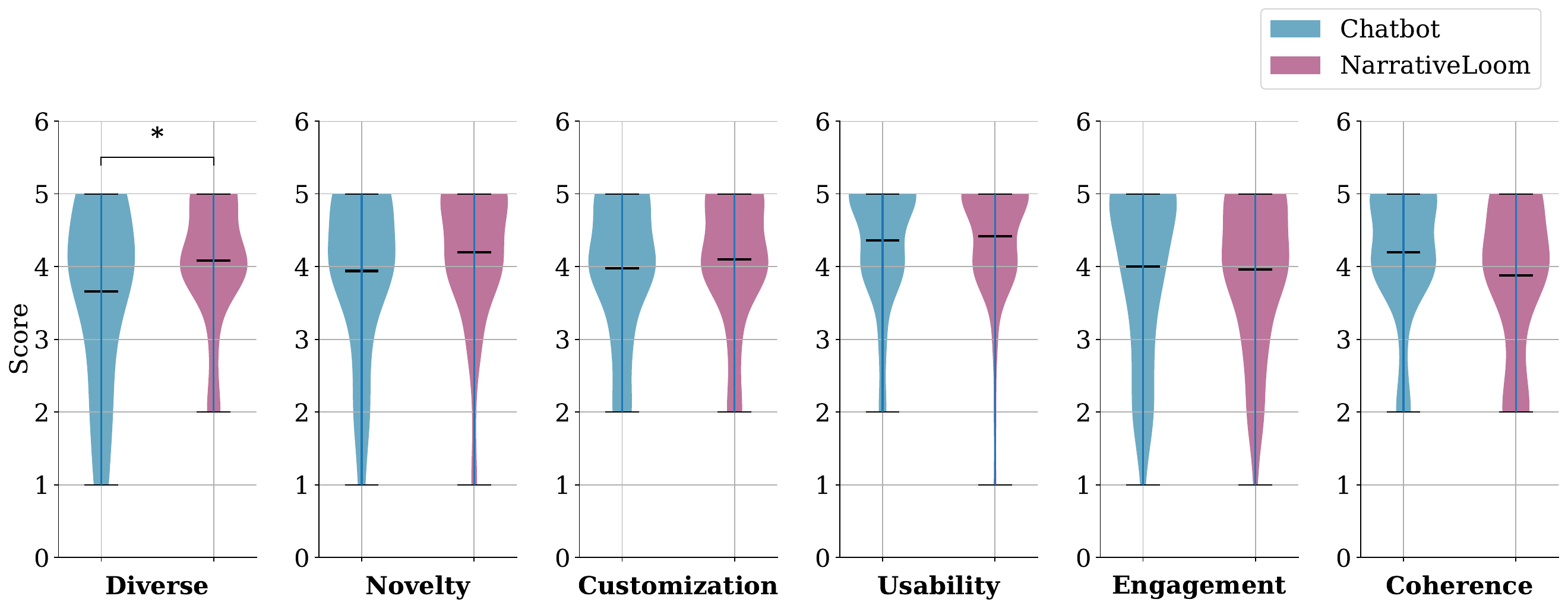}
    \caption{\textbf{The comparison of user evaluation scores between \model and the chatbot.} Violin plots show the distribution of user ratings on a 5-point scale across key dimensions. \model consistently achieved higher median scores for creativity-focused metrics like diversity and novelty, while performing comparably on usability, engagement, and coherence. Asterisks denote statistical significance: * indicates $p$ < 0.05.}
    \Description{A series of six violin plots arranged horizontally, comparing two systems (baseline chatbot in blue and \model in pink) across six dimensions: Novelty, Diversity, Coherence, Customization, Engagement, and Usability. Each violin plot shows the distribution of user ratings from 1 to 5 on the y-axis. The pink violins (\model) show higher median scores for Novelty and Diversity, with an asterisk above Diversity indicating statistical significance. For Coherence, the blue violin shows a slightly higher median. For Customization, Engagement, and Usability, the two distributions overlap considerably with similar medians around 4.0. The violin widths indicate the density of responses at each rating level.}
    \label{fig:user_comparison} 
\end{figure*}

\section{Findings}

\subsection{Strategic Exploration of a Diverse Narrative Landscape}
\label{f1}

To evaluate whether \model successfully enhanced creative exploration through diverse narrative possibilities, we analyzed both user perceptions of creativity and the underlying patterns of how users strategically engaged with the multi-persona system.

\paragraph{User-Perceived Creative Enhancement}

In terms of user-perceived creative enhancement, \model demonstrated a practically meaningful advantage over the baseline (see \cref{fig:user_comparison}). Analysis using a paired samples t-test confirmed that the improvement in \textbf{diversity} (see \cref{metric}) was statistically significant ($M=4.08$, $SD=0.89$ \vs $M=3.66$, $SD=1.23$; $t(49)=2.14$, $p=0.037$, Cohen's $d=0.39$). For \textbf{novelty} (see \cref{metric}), while the higher ratings ($M=4.20$, $SD=0.98$ \vs $M=3.94$, $SD=1.10$; Cohen's $d=0.25$) did not reach statistical significance, the effect size still indicated a trend favoring our system.

Users consistently praised \model's creative diversity, with P23, P24, and P34 noting it offered ``\textit{much more choice of plot direction}'' and generated ideas that ``\textit{seemed more diverse and novel.}'' Users particularly valued the structured approach to creative exploration, with P9 explaining that \model \textit{``allows having the different beats added layers to the story instead of, in a chatbot, creating just one part of the story.''} .

\paragraph{Frequency and Narrative Roles.}

\begin{figure*}[htbp]
    \centering
    \begin{subfigure}{0.49\linewidth}
        \centering
        \includegraphics[width=\linewidth]{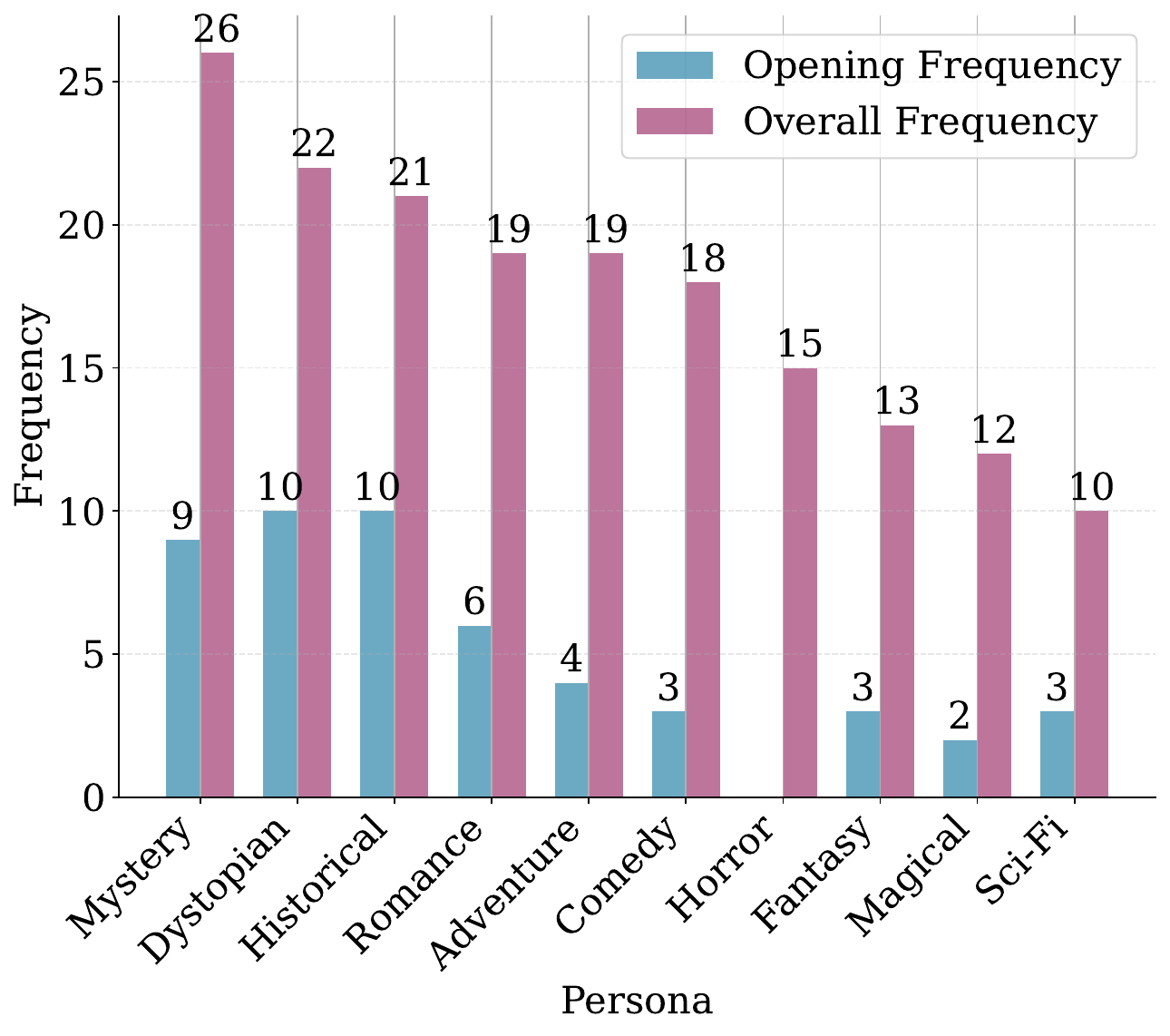}
        \caption{\textbf{Persona usage roles.} The grouped bar chart reveals specialized roles. Personas like \textbf{Historical} and \textbf{Dystopian} were primary ``initiators'' with high opening usage, while \textbf{Mystery} was the key ``developer,'' dominating overall usage.}
        \label{fig:persona_frequencies}
    \end{subfigure}
    \hfill 
    \begin{subfigure}{0.49\linewidth}
        \centering
        \includegraphics[width=\linewidth]{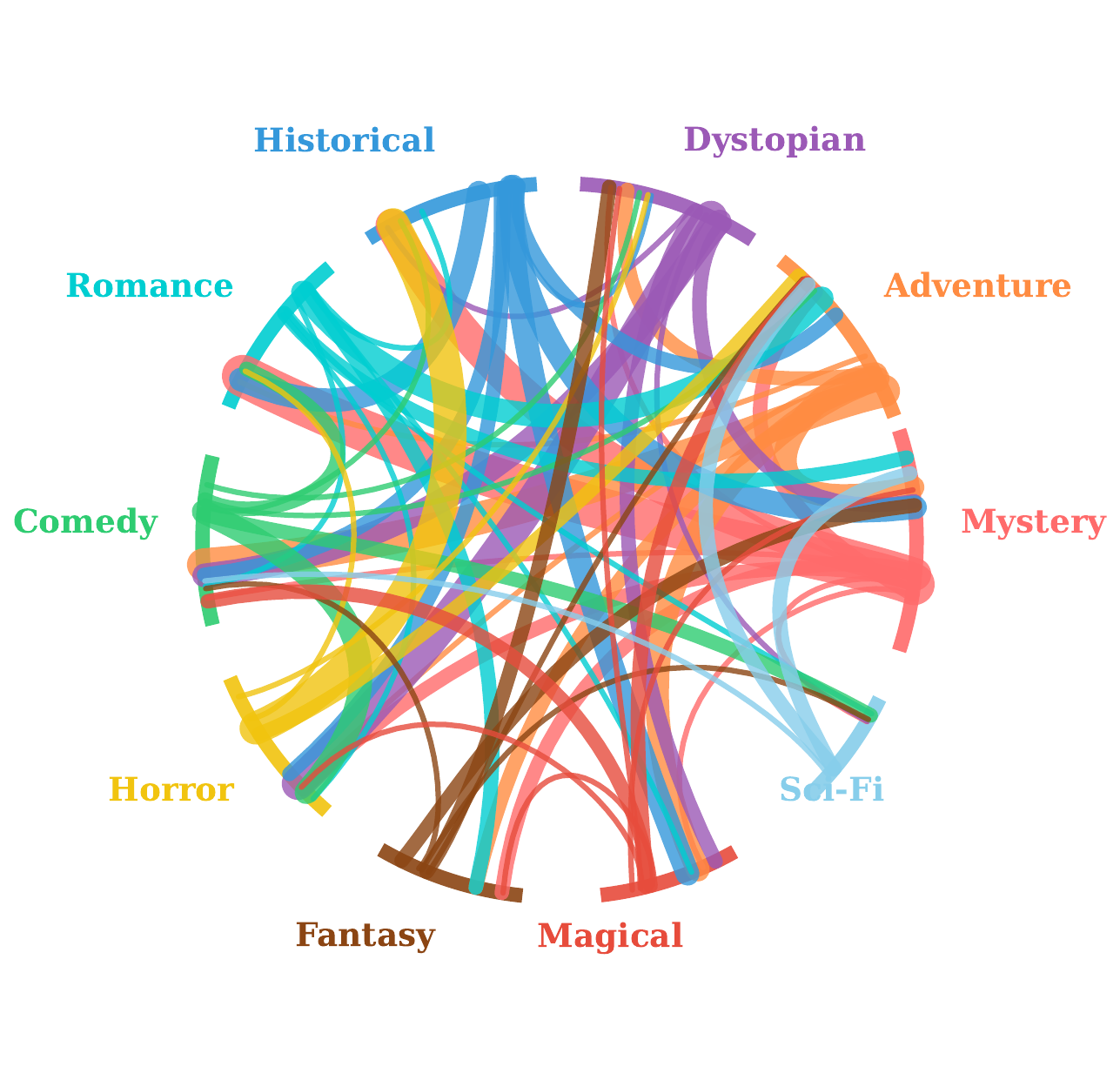}
        \caption{\textbf{Narrative flow network.} The transition graph shows that movement between personas was not random. Edge thickness indicates frequency, revealing popular, often asymmetric pathways (\eg, from Mystery to Romance).}
        \label{fig:transition_network}
    \end{subfigure}
    \caption{\textbf{Users strategically selected personas and transitioned between them in structured patterns.} The data reveals two key user behaviors: (a) selecting personas for specialized ``initiator'' or ``developer'' roles, and (b) moving between personas along logical, genre-adjacent pathways.}
    \Description{Two visualizations side by side. Left panel: A grouped bar chart with ten persona types on the x-axis (Fantasy, Sci-Fi, Mystery, Romance, Historical, Horror, Adventure, Comedy, Dystopian, Magical) and frequency counts on the y-axis from 0 to 50. Each persona has two bars: a lighter bar showing opening usage and a darker bar showing overall usage. Historical and Dystopian show tall light bars (high opening usage) while Mystery shows the tallest dark bar (high overall usage). Right panel: A circular chord diagram with ten personas arranged around the perimeter. Curved ribbons connect different personas, with ribbon thickness indicating transition frequency. The thickest ribbons connect Mystery to Romance, Adventure to Comedy, and Dystopian to Horror, showing preferred transition pathways between genres.}
    \label{fig:persona_interaction_patterns}
\end{figure*}

An analysis of persona usage patterns revealed that users strategically selected specific personas for specialized roles depending on the narrative stage (see \cref{fig:persona_frequencies}). While the Historical and Dystopian personas were the most frequent choices for story initiation, the Mystery persona maintained the highest overall usage frequency. This contrast suggests an intuitive functional assignment by users: personas with strong world-building archetypes (Historical and Dystopian) served as ``initiators,'' establishing the setting and initial conflict, whereas the Mystery persona emerged as the primary ``developer,'' utilized to navigate plot complexities and advance the narrative once the story was underway.

\paragraph{Transitions and Narrative Flow.}

To understand how stories evolved, we analyzed the transition network between personas (see \cref{fig:transition_network}). The transitions were not random but formed a structured network characterized by high-frequency pathways and strong directional preferences. The flow from Mystery to Romance was the most common, while other prominent paths included Adventure to Comedy and Dystopian to Horror. These pathways often represent complementary genre pairings, suggesting users were actively blending genres to create richer narratives. Furthermore, this flow was significantly asymmetric, with the average difference between forward and reverse transitions being statistically significant ($M = 1.24 \pm 0.92$, $t(45) = 8.95$, $p < 0.001$). For example, the transition from Mystery to Romance (5 instances) was more than twice as common as the reverse (2 instances), while the Historical to Magical path (3 instances) was never reciprocated. This directionality demonstrates that users guided their stories along specific trajectories, leveraging the multi-persona system not for simple variety, but as a toolkit for purposeful, structured narrative escalation.

\subsection{Balancing Co-Creative Agency with Scaffolding}
\label{f2}

Quantitatively, \model performed comparably to the baseline across three dimensions (see \cref{metric}): \textbf{customization} (\model: $M=4.10, SD=0.92$ \vs chatbot: $M=3.98, SD=0.97$), \textbf{usability} (\model: $M=4.42, SD=0.83$ \vs chatbot: $M=4.36, SD=0.84$), and \textbf{engagement} (\model: $M=4.00, SD=1.15$ \vs chatbot: $M=3.96, SD=1.04$), with no statistically significant differences. This suggests that managing multiple personas did not compromise usability despite the added complexity. Beyond these quantitative similarities, our qualitative analysis reveals distinct forms of creative control enabled by \model. We identify three key dimensions of enhanced co-agency.

\paragraph{Distributed Creative Authority (Shared Agency)} 

\model's multi-persona architecture distributed creative authority between system and user. Rather than producing a single linear suggestion, the personas generated diverse narrative directions that expanded the creative space. Users remained in charge of selecting, combining, or discarding these options, thereby maintaining clear editorial control. Participants valued this balance of shared agency; for instance, P23 remarked, ``\textit{I liked the different options from the agents},'' while P24 emphasized that it offered ``\textit{much more choice of plot direction}'' compared to a traditional chatbot that typically steers the story along a single path.  

\paragraph{Responsive Collaborative Refinement (Negotiated Agency).} 

Users experienced agency as a negotiated process where their feedback actively redirected the system’s output. Multiple participants emphasized the system’s ability to integrate their feedback into subsequent generations. For example, P19 noted that it ``\textit{took into account edits and comments alongside coming up with engaging scenarios and characters}.'' This iterative responsiveness illustrates how user input and system creativity co-shaped the evolving trajectory of the story.  

\paragraph{Empowered Creative Partnership (Augmented Agency).} 

By amplifying users’ expressive capacities, \model shifted the experience toward an empowered sense of agency. Participants described how the system ``\textit{generated a comprehensive and creative story using my prompts (P30)}'' and produced narratives that ``\textit{read so much more naturally and fuller compared to the others (P24)}.'' P45 reflected this heightened agency through strong engagement: ``\textit{I love the different stories the system generated for me. It was captivating and really interesting}.''  

\subsection{Navigating the Creativity-Coherence Trade-off with a Segmented Structure}
\label{f3}

\begin{figure*}[htbp]
    \centering
    \includegraphics[width=\linewidth]{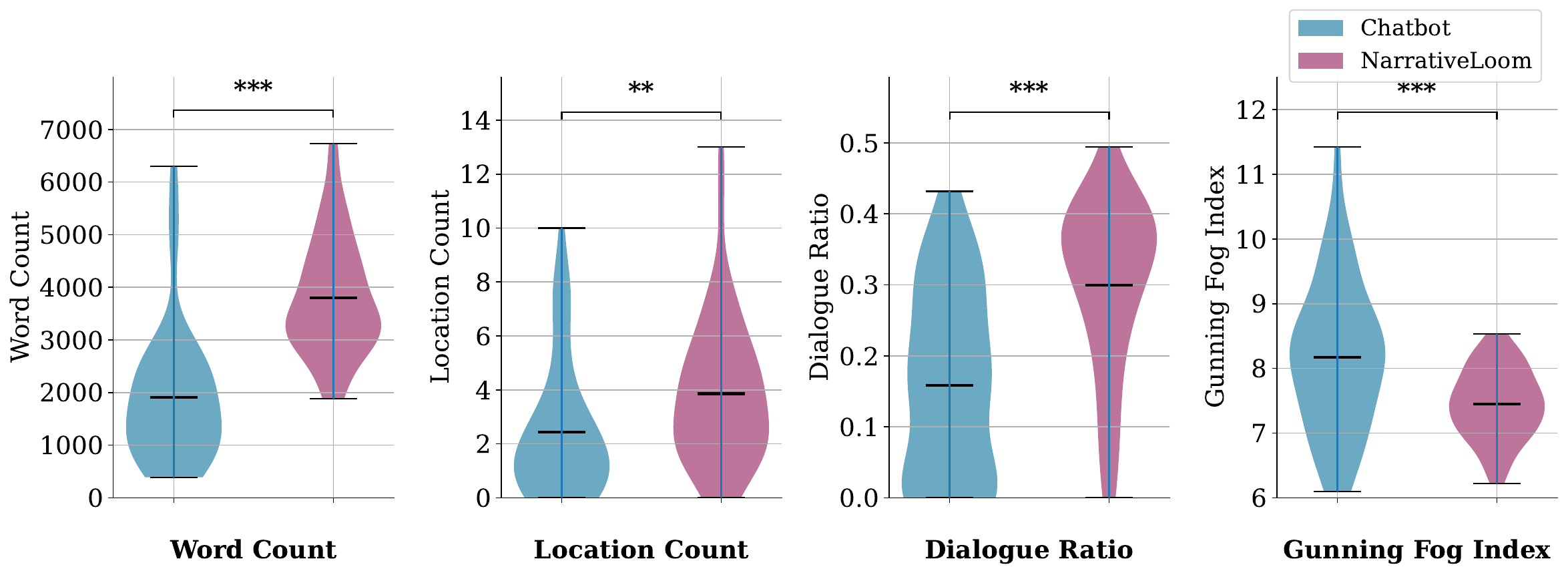}
    \caption{\textbf{The comparison of narrative features between \model and the chatbot.} Violin plots compare the chatbot baseline (blue) with \model (pink). \model generated stories that were significantly longer, contained more locations, had a higher ratio of dialogue, and were more readable (lower Gunning Fog Index). Asterisks denote statistical significance: ** indicates $p$ < 0.01 and *** indicates $p$ < 0.001.}
    \Description{Four violin plots arranged horizontally comparing two systems across narrative metrics. First plot (Word Count): Pink violin extends from approximately 2000 to 6000 words with median around 3800, while blue violin ranges from 0 to 4000 with median around 1900, marked with three asterisks. Second plot (Location Count): Pink violin ranges 0--10 with median around 4, blue violin ranges 0--8 with median around 2, marked with two asterisks. Third plot (Dialogue Ratio): Pink violin ranges 0.0--0.5 with median around 0.30, blue violin ranges 0.0--0.4 with median around 0.16, marked with three asterisks. Fourth plot (Gunning Fog Index): Pink violin ranges 6.5--8.5 with median around 7.5, blue violin ranges 6.0--11.0 with median around 8.2, marked with three asterisks. All pink distributions show more favorable outcomes.}
    \label{fig:narrative_features_comparison} 
\end{figure*}

\model's structured approach, which breaks the narrative into manageable creative units, yielded significant advantages in story development and richness (see \cref{fig:narrative_features_comparison}). \model enabled users to create substantially \textbf{longer} stories ($M=3803.16,SD=1109.50$ words) than the chatbot baseline ($M=1907.88,SD=1304.44$ words; $t(49)=9.160, p<.001$). This structure also supported the creation of richer narrative environments, with stories from \model incorporating significantly more \textbf{locations }($M=3.86,SD=2.67$ \vs $M=2.44,SD=2.54$; $t(49)=3.276, p=.002$) and a higher ratio of \textbf{dialogue} ($M=0.30,SD=0.12$ \vs $M=0.16,SD=0.13$; $t(49)=5.675, p<.001$). Furthermore, the resulting narratives were more accessible, achieving a lower (better) \textbf{readability} score ($M=7.45,SD=0.54$ \vs $M=8.17,SD=1.16$; $t(49)=-4.212, p<.001$).

The qualitative feedback confirmed the value of this segmented process. As P10 noted, ``I liked how \model approached the storytelling process in sections. chatbot felt more all or none while the system worked on one moment at a time.'' However, this same multi-persona, segmented architecture that fostered creativity also introduced the challenge of maintaining narrative coherence.

Indeed, the chatbot baseline achieved a higher \textbf{coherence} rating (see \cref{fig:user_comparison}; $M=4.20,SD=0.92$) compared to \model ($M=3.88,SD=1.05$).While this difference was not statistically significant, it represented a small-to-medium effect size (Cohen's $d=0.32$). This is an expected trade-off: the chatbot generates a continuous, single-voice narrative, while our system's multi-persona design and significantly longer stories create more potential fragmentation points where consistency can be challenged.

Despite these results, users reported that \model maintained logical consistency. P28 noted that ``overall the suggestions and generated text were good and the process of generating the story was smooth,'' while P38 observed that the system ``created compelling chunks that followed through to create a fluid story.'' These qualitative reports suggest that the system manages the coherence-creativity trade-off, increasing narrative richness while maintaining acceptable story coherence.

\begin{figure*}[htbp]
    \centering
    \includegraphics[width=\linewidth]{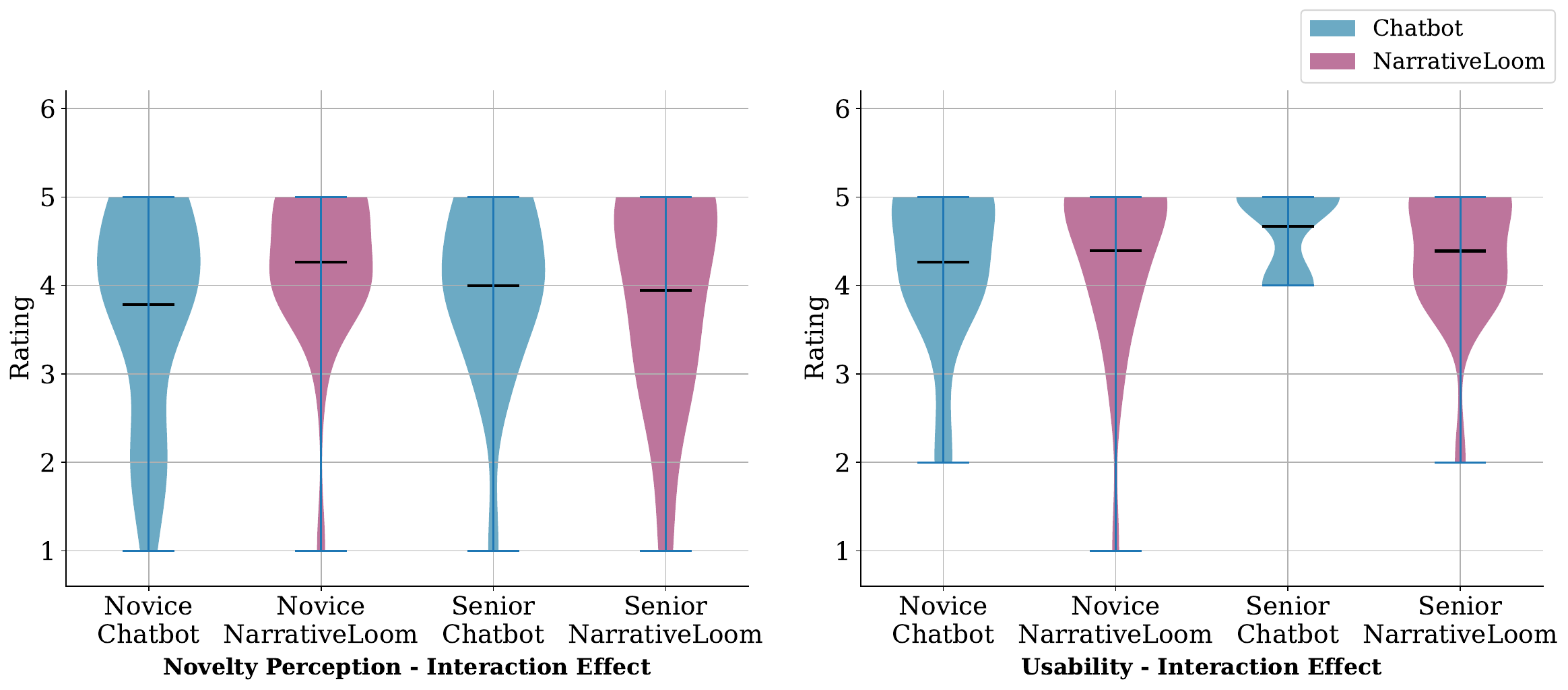}
    \caption{\textbf{Interaction between writing expertise and system preference.} Violin plots show user ratings for novelty perception (left) and usability (right), comparing \model (purple) against the chatbot baseline (blue). Novice writers perceived \model as more novel, whereas senior writers rated the chatbot higher on usability.}
    \Description{Two sets of violin plots showing expertise-based differences. Left panel (Novelty Perception): For novice writers, purple violin (\model) shows higher median around 4.3 compared to blue violin (chatbot) at 3.8. For senior writers, distributions are more similar with medians around 4.0. Right panel (Usability): For novice writers, both systems show similar distributions with medians around 4.3. For senior writers, blue violin shows higher median around 4.7 while purple violin shows median around 4.4. Each violin's width indicates the concentration of ratings, with all ratings on a 1-5 scale on the y-axis.}
    \label{fig:interaction_effect}
\end{figure*}

\subsection{Writing Expertise Shapes Preferences for AI Writing Assistance}
\label{f4}

User characteristics, specifically writing experience, influenced interactions with the storytelling systems, suggesting a need for personalized \ac{ai} writing assistance. We observed different patterns in how novice and senior writers evaluated the two systems (see \cref{fig:interaction_effect}), although the interaction effects were not statistically significant.

Regarding novelty, novice writers rated \model higher than the chatbot baseline (\model: $M=4.26, SD=0.90$; chatbot: $M=3.78, SD=1.18$; Cohen's $d=0.457$). Conversely, senior writers showed no clear preference, rating both systems similarly (chatbot: $M=4.00, SD=1.00$; \model: $M=3.94, SD=1.18$). 

Usability ratings followed a different pattern. Novices reported a slight preference for \model (\model: $M=4.39, SD=0.97$; chatbot: $M=4.26, SD=0.90$; Cohen's $d=0.140$), whereas senior writers rated the chatbot's usability higher (chatbot: $M=4.67, SD=0.47$; \model: $M=4.39, SD=0.76$; Cohen's $d=0.441$). 

User preferences were also influenced by their specific writing stage. P8 summarized this distinction: \textit{``Chatbot is suitable to one who knows their rough story already and just needs help making it. But the \model system for those who want to write but are not sure what story to tell.''} This suggests that \ac{ai} assistance should align with different phases of the creative process. The structured \ac{bvsr} framework in \model supports the ideation and exploration phase by helping writers identify narrative possibilities. In contrast, traditional chatbots may be more effective during the development phase, when writers focus on executing and refining established concepts.

These results indicate that a universal approach to \ac{ai} writing assistance may be suboptimal. The observed preference patterns---novices prioritizing the novelty of \model and senior writers prioritizing chatbot usability---suggest that support systems should be tailored to both user expertise and their current stage in the writing process.

\subsection{Expert Evaluation Results}
\label{f5}

To complement the large-scale user study, four creative writing experts assessed narrative quality. We randomly sampled 20 story pairs from the dataset, each consisting of one story generated by \model and one by the chatbot baseline ($N=40$). The experts demonstrated a near-unanimous preference for \model. While noting that neither system consistently produced perfect outputs, the experts rated the \model narratives as higher quality and more compelling than those from the baseline.

\paragraph{Quantitative Preference and Creativity Scores} 

\begin{figure*}[htbp]
    \centering
    \includegraphics[width=\linewidth]{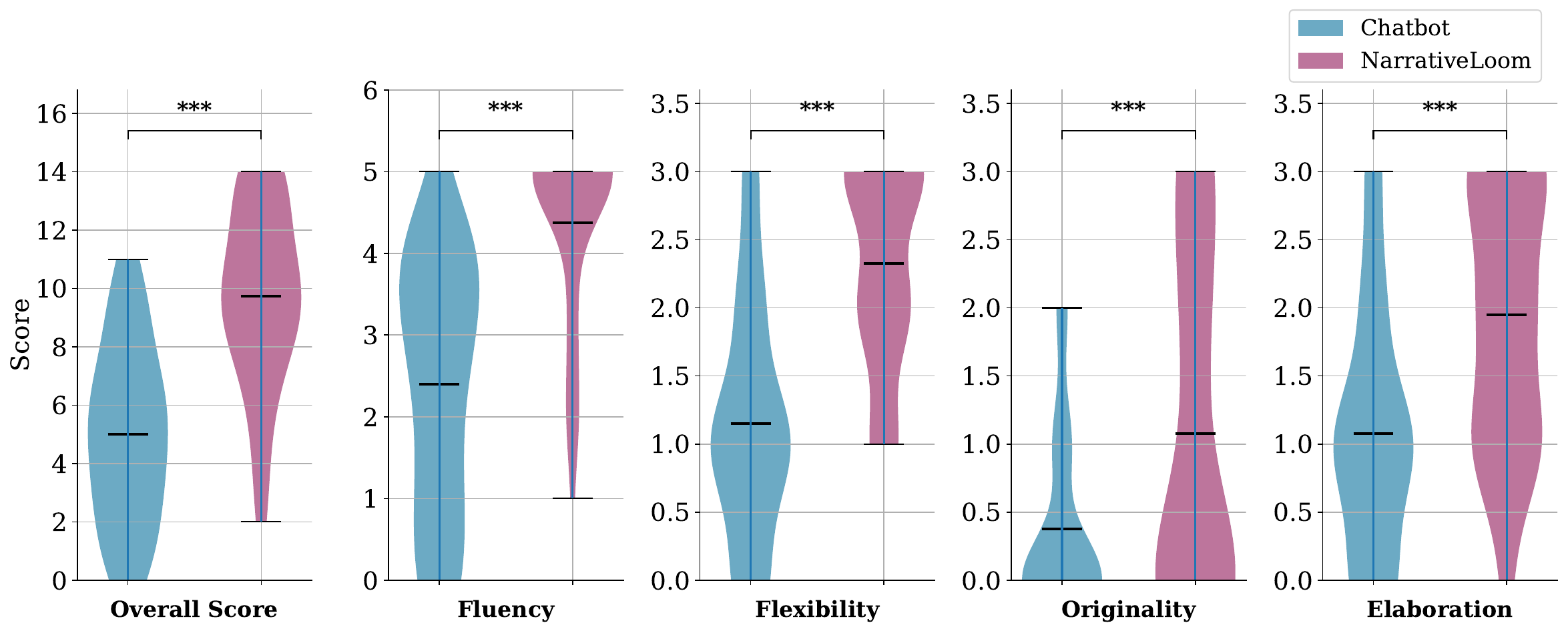}
    \caption{\textbf{Expert evaluation results using the \ac{ttcw} framework.} Violin plots show the distribution of creativity scores across four dimensions and overall performance, comparing \model (purple) against the chatbot baseline (blue). \model demonstrated statistically significant improvements across all creativity dimensions. Asterisks denote statistical significance: *** indicates $p$ < 0.001.}
    \Description{Five violin plots arranged horizontally showing expert ratings. First plot (Overall Score): Purple violin ranges 4--14 (out of 14) with median around 10, blue violin ranges 2--9 with median around 5. Second plot (Fluency, out of 5): Purple median around 4.5, blue median around 2.5. Third plot (Flexibility, out of 3): Purple median around 2.3, blue median around 1.2. Fourth plot (Originality, out of 3): Purple median around 1.1, blue median around 0.4. Fifth plot (Elaboration, out of 3): Purple median around 2.0, blue median around 1.1. All comparisons are marked with three asterisks indicating strong statistical significance. Purple distributions consistently show higher scores and less variance than blue distributions.}
    \label{fig:expert_result}
\end{figure*}

\begin{figure*}[htbp]
   \centering
   \includegraphics[width=\linewidth]{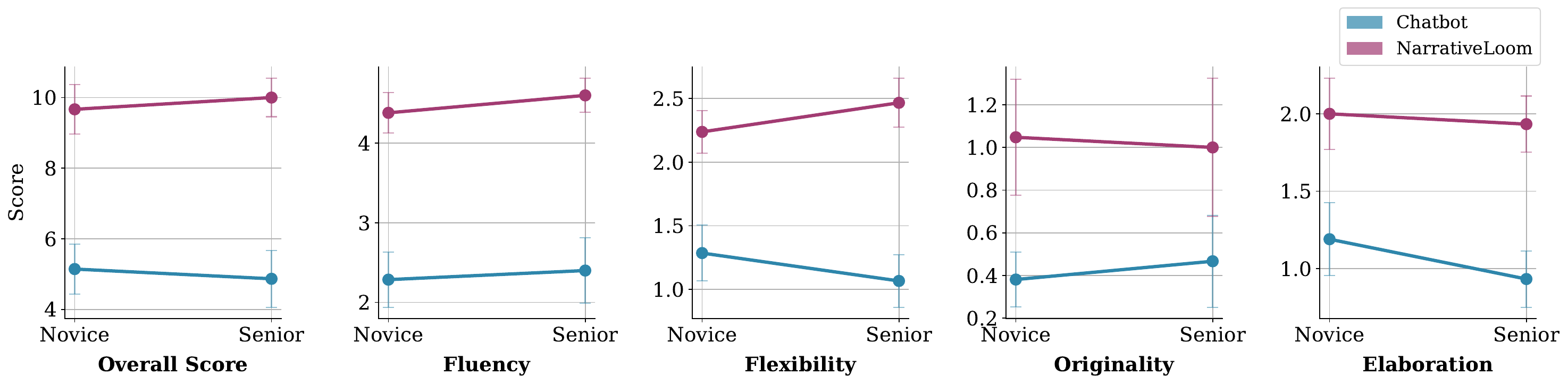}
   \caption{\textbf{Expert \ac{ttcw} evaluations participant narratives based on writing expertise.} Mean scores ($\pm SE$) are shown for novice and senior writers using \model (purple) and chatbot (blue). Evaluation criteria include Overall Score, Fluency, Flexibility, Originality, and Elaboration. \model outperformed the chatbot across all metrics for both user groups.}
   \Description{Five interaction plots arranged horizontally showing mean scores with error bars. Each chart compares four conditions: novice with chatbot (blue), novice with \model (purple), senior with chatbot (blue), and senior with \model (purple). First chart (Overall Score, out of 14): Purple line shows means around 9-10 for both expertise levels, blue line shows means around 5 for both levels. Second chart (Fluency, out of 5): Purple line around 4.3-4.5, blue line around 2.3-2.5. Third chart (Flexibility, out of 3): Purple line around 2.2-2.4, blue line around 1.1-1.2. Fourth chart (Originality, out of 3): Purple line around 1.0-1.1, blue line around 0.3-0.5. Fifth chart (Elaboration, out of 3): Purple line around 1.9-2.0, blue line around 1.0-1.1. Across all dimensions, purple lines are consistently taller than blue lines, and differences between novice and senior within the same system are minimal.}
   \label{fig:expertise_ttcw_plot}
\end{figure*}

Experts preferred \model\ in direct comparisons, selecting stories generated by \model\ in 38 out of 40 forced-choice scenarios. This preference is reflected in the \ac{ttcw} scores (\cref{fig:expert_result}). A paired t-test showed that \model\ achieved higher overall creativity scores ($M = 9.72$ out of 14, $SD = 2.1$) than the chatbot baseline ($M = 5.00$ out of 14, $SD = 1.8$; $t(79) = 13.68, p < 0.001$). Analysis of individual dimensions showed advantages across fluency ($M = 4.38$ out of 5 \vs\ $2.40$; $t(79) = 12.93, p < 0.001$), flexibility ($M = 2.33$ out of 3 \vs\ $1.15$; $t(79) = 12.61, p < 0.001$), elaboration ($M = 1.95$ out of 3 \vs\ $1.07$; $t(79) = 8.31, p < 0.001$), and originality ($M = 1.07$ out of 3 \vs\ $0.38$; $t(79) = 6.16, p = 0.001$). Furthermore, \model\ outperformed the chatbot baseline across all user expertise levels; the magnitude of this advantage did not differ significantly between novice and senior writers (\cref{fig:expertise_ttcw_plot}). Although the small expert sample size limits broad statistical inference, these results align with our larger-scale user study.

\paragraph{Creative Originality and Unpredictability}

Expert qualitative feedback identified three primary areas where \model\ produced superior narratives, centering first on creative innovation. The baseline was frequently criticized for relying on convention; E1 noted that it \textit{``frequently relies on conventional story structures and clichéd elements.''} In contrast, \model was lauded for its narrative surprise. As E4 remarked, \textit{``These stories take you to a place that you don't expect... The chatbot stories are kind of predictable.''} This was echoed by E3, who noted, \textit{``This story generated by \model is a bit of a surprise. I didn't expect it when reading the [sparkle].''} Experts also pointed to specific novel ideas, with E1 highlighting that \textit{``stories from \model often have some clever ideas or highlights, things that I haven't seen before and are less clichéd. For example, for the `friendship' topic, a time traveler appears after the friends meet.''}

\paragraph{Superior Narrative Craft: ``Showing'' Over ``Telling''}

A second theme involved \model's use of immersive, scene-based storytelling. E1 contrasted the two systems by describing that the chatbot \textit{``is a bit like an instruction manual, laying things out directly from a 'God's eye view' and simply listing events.''} In contrast, E1 observed that \textit{``\model doesn't present all the information at once; instead, it starts with a scene, giving the feeling of a story being slowly and engagingly told.''} This aligns with E2's comment that \model's writing is \textit{``deeply internal, rich with details and expanded scenic development,''} whereas the chatbot is \textit{``more third-person, objective, and summary-like.''} The distinction was articulated most directly by E4: \textit{``\model has a much higher level in the way it is written. It is better at showing not telling.''}

\paragraph{Character Depth and Psychological Realism}

Finally, experts noted \model's ability to render characters with psychological depth, a quality they found lacking in the baseline. E1 observed that \textit{``most characters in \model's stories are well set,''} and E4 summarizing the preference directly by stating that \model \textit{``has better characters and dialogues.''} E2 elaborated on this distinction, noting that chatbot characters \textit{``lacked internal depth and psychological complexity with no inner hesitation or psychological processing.''} Conversely, \model\ produced more nuanced figures; E2 highlighted \model's \textit{``strong character development,''} particularly regarding antagonists who felt like \textit{``real people rather than artificial constructs.''} This capacity for believable interiority was a consistent differentiator across the expert reviews.

\section{Discussion}

Our study showed that \model significantly improved creative storytelling compared to the chatbot system (see \cref{f5}), with users strategically leveraging different personas for specialized narrative roles (see \cref{f1}), producing longer, richer-detailed stories (see \cref{f3}), and maintaining creative agency despite the system's structured guidance (see \cref{f2}). These findings illuminate how Campbell's \acp{bvsr} theory~\citep{campbell1960blind} can serve as a principled framework for designing \ac{ai} systems that enhance creative possibilities while preserving user agency.

\subsection{Balancing Creative Diversity with Cognitive Manageability}

Our findings reveal a fundamental tension in AI-assisted creative writing: how to expand narrative possibilities while maintaining cognitive manageability. Current systems address this through interventions at different narrative layers. Systems operating at the \textbf{discursive layer} through parametric sampling (Wordcraft~\citep{yuan2022wordcraft}) enable stylistic variation but cannot alter underlying story logic. The \textbf{emotional trajectory layer} (TaleBrush~\citep{chung2022talebrush}) provides visual story arc control but constrains narrative specificity. The \textbf{structural planning layer} offers either rigid top-down cascading (Dramatron~\citep{mirowski2023co}) that limits spontaneity or configurable abstraction (WhatElse~\citep{lu2025whatelse}) that demands complex pre-planning.

Our genre-based persona approach intervenes at the \textbf{diegetic layer}---the level of story events and causal logic. This distinction matters because genres shape narrative possibility spaces~\citep{chandler1997genre}. By instantiating multiple genre personas, we address the statistical centroid problem~\citep{bender2021dangers}, where single models converge on predictable outputs by sampling from the center of their training distribution. Each persona samples from distinct narrative regions: when participants used comedy for tension relief or mystery for plot complexity, they accessed fundamentally different narrative logics. This explains why experts found \model stories more likely to take readers to unexpected places while baseline stories remained predictable---genre intervention alters \textit{what can happen}, not just \textit{how it's told}.

The beat-based design provides a specific granularity for creative scaffolding. Dramatron's~\citep{mirowski2023co} rigid planning ensures coherence but demands heavy cognitive pre-planning. Wordcraft's~\citep{yuan2022wordcraft} open-ended generation offers flexibility but can overwhelm users. Our approach provides a middle-out position---semantic anchors that enable structured improvisation without predetermined outcomes. This operates within the zone of proximal development~\citep{vygotsky1978mind}: it is complex enough for sophisticated narratives yet modular enough to remain manageable. The significantly longer stories, richer dialogue, and diverse settings in \model validate this balance. Minor coherence trade-offs represent acceptable costs for enhanced creative exploration.

\subsection{Temporal Synthesis and Collaborative Emergence}

The locus of creativity in \ac{bvsr} remains a subject of ongoing debate. While classical theory identifies variation as the primary creative driver---\citet{campbell1960blind} and \citet{simonton1999creativity,simonton2023blind} argue that creativity arises from numerous blind variations---recent computational research suggests that creative agency resides in human selection. In this view, internal predictive models guide strategic choices rather than stochastic filtering~\citep{dietrich2015human,zhu2017creativity}.

Our results indicate that creativity in \model emerges from a triadic interaction between algorithmic variation, human selection, and contextual synthesis. Persona transition patterns (see \cref{fig:transition_network}) show that users do not merely select options; they construct narrative trajectories through sequential choices, where each selection constrains subsequent possibilities. Asymmetric transitions suggest that users exploit narrative affordances~\citep{turner2025narrative} arising from the intersection of multiple personas. This observation aligns with \citet{zhou2024generative}, who noted in text-to-image contexts that human creativity manifests through curation. It further supports \citet{epstein2023art}'s argument that generative \ac{ai} shifts creativity from direct manipulation to iterative specification, allowing users to maintain control over the narrative trajectory.

However, our findings identify a distinction relevant to \ac{hci} design: unlike image generation~\citep{zhou2024generative}, which involves selecting from parallel alternatives, narrative creation requires temporal synthesis. Users must integrate selected beats into coherent sequential structures. This temporal dimension facilitates collaborative emergence~\citep{sawyer2000improvisation}, where creative products arise from structured improvisation between human and \ac{ai}. The higher dialogue ratios and increased location diversity in \model stories suggest that this synthesis amplifies creative elaboration beyond individual capabilities, supporting theories of distributed cognition \citep{hutchins2000distributed}. These results suggest that co-creative systems should function as platforms for temporal synthesis rather than simple generation tools, leveraging both human curation and \ac{ai} variation.

\subsection{Writers Require Varying Degrees of Creative Support}

Our findings reveal that writers require different types of creative support based on their \textbf{expertise level} and \textbf{creative stage} (see \cref{f4}), challenging the one-size-fits-all approach in current \ac{ai} writing tools. This aligns with prior work suggesting that writers seek diverse roles from \ac{ai} collaborators~\citep{chakrabarty2025can,gero2023social}, yet illuminates a critical design consideration often overlooked in creativity support research~\citep{cherry2014quantifying}.

Consistent with expert-novice differences in creative domains~\citep{ericsson2018differential,ericsson1991toward}, we observed separate preference patterns. Novice writers favored \model's structured exploration, benefiting from multi-persona scaffolding that activates different narrative spaces~\citep{sio2015fixation} and helps overcome creative blocks when their own resources are limited~\citep{yuan2022examples,weisberg2006modes}. Conversely, expert writers preferred streamlined, chatbot-like interfaces that integrate seamlessly into their established workflows. This aligns with the expertise reversal effect~\citep{kalyuga2009expertise}---scaffolding beneficial to novices becomes extraneous cognitive load for experts who possess highly developed schemas and seek to maintain creative flow~\citep{csikszentmihalyi1997flow}.

Crucially, our \ac{ttcw} evaluation revealed that these preference differences did not translate into quality differences---both groups achieved comparable creative improvements with \model. This paradox illuminates a key distinction: creative support operates at two levels---\textbf{workflow integration} (where experts favor minimal disruption) and \textbf{cognitive stimulation} (where structured diversity benefits all). The multi-persona system's value lies in combating cognitive fixation through systematic variation, a benefit that persists regardless of expertise level.

Beyond static expertise, support needs vary dynamically by creative phase. Writers may benefit from \model's exploratory scaffolding during ideation but prefer streamlined assistance during revision~\citep{finke1996creative,chakrabarty2024creativity,ippolito2022creative}. This suggests future creative \ac{ai} should implement \textbf{adaptive dual-channel designs}: preserving cognitive benefits of structured variation while allowing interface customization from fully scaffolded (novices/ideation) to minimally intrusive (experts/refinement). The key principle is decoupling creative scaffolding from interface complexity---maintaining the former's value while minimizing the latter's disruption.

\subsection{Design Implications for Creative AI Systems}

\subsubsection{Theory-Guided Design as a Counterpoint to Model Scale.} 

Our results show that creative \ac{ai} systems benefit from deliberate, theory-guided frameworks rather than relying solely on increased parameter scale to improve performance~\citep{brown2020language,peng2025probing}. The success of our structured multi-persona model, grounded in Campbell’s \ac{bvsr} framework~\citep{campbell1960blind}, suggests that principled design can be more effective than raw computational power. This aligns with arguments that current \ac{ai} approaches may face diminishing returns without fundamental structural innovation~\citep{marcus2020next}. Instead of pursuing larger models, our approach demonstrates that implementing diverse creative perspectives allows systems to explore narrative spaces often underrepresented in training corpora, such as those requiring novel genre combinations or unconventional plot developments. Future creative \ac{ai} could employ modular systems with specialized components optimized for specific creative functions, providing more targeted assistance.

\subsubsection{Structured Improvisation Architecture for Creative Emergence.} 

The beat-based segmentation approach addresses a core challenge in long-form generation: maintaining coherence without stifling emergence. Unlike hierarchical planning systems that may constrain discovery~\citep{mirowski2023co} or unconstrained generation that risks incoherence~\citep{holtzman2020curious}, our model enables progressive coherence---local consistency within manageable units paired with emergent global development. These findings suggest that creative \ac{ai} should decompose complex tasks into discrete, context-aware units that preserve spontaneity while providing a navigational structure for extended development~\citep{sawyer2014group}. Future systems could use adaptive mechanisms to automatically identify optimal boundaries based on project type, user preference, and context, scaling from short-form to extended works while maintaining both novelty and logic.

\subsubsection{Explicit Role Separation for Creative Agency.} 

The balance between \ac{ai} assistance and human ownership in \model highlights the necessity of explicit role separation. By assigning \ac{ai} the responsibility for variation generation and reserving selection authority for the human, our approach maintained user agency while leveraging computational speed. Creative \ac{ai} systems should therefore implement transparent boundaries between machine and human contributions, avoiding black box patterns that undermine creator confidence~\citep{zhu2018explainable,kantosalo2016modes,davis2015enactive}. Future designs should include explicit role indicators, separate interfaces for generation and selection, and allow users to negotiate these boundaries by choosing how much creative territory to delegate.

\subsection{Limitations and Future Work}

While our results demonstrate the effectiveness of \ac{bvsr}-based computational creativity, several theoretical and methodological limitations warrant consideration. The reliance on genre-based persona specialization, while effective for narrative diversity, may tend to draw upon established literary conventions rather than fully exploring unconventional creative possibilities~\citep{chandler1997genre}. Our evaluation framework, though comprehensive, necessarily reflects Western narrative traditions and assessment criteria. Cross-cultural validation would be essential to establish broader applicability, particularly given evidence that creativity manifestation and evaluation vary significantly across cultural contexts~\citep{niu2001cultural}. Additionally, the observed expertise effects raise questions about the long-term developmental implications of AI-assisted creativity. While novices benefited from system support, the long-term developmental impact remains unclear. Longitudinal studies are necessary to determine whether such scaffolding facilitates skill acquisition or inadvertently creates dependency, potentially impeding the development of independent creative capabilities. This concern mirrors established debates regarding technological scaffolding in educational contexts~\citep{pea2004social}, where over-reliance on external support can diminish the internal cognitive effort required for mastery. Investigating this trajectory is vital for designing systems that serve as developmental tools rather than mere cognitive crutches.

Future work will prioritize four system enhancements. First, we will \textbf{extend beyond genre-based specialization through function-oriented modules}, developing personas focused on world-building, character psychology, and narrative pacing. This hybrid system aims to enable unconventional creative combinations while retaining the genre diversity benefits observed in our study. Second, we plan to \textbf{implement adaptive weighting mechanisms} that learn from user selection patterns to adjust persona prominence. By proactively suggesting complementary alternatives whenever the system senses a decline in creative divergence, this approach seeks to mitigate creative fixation. Third, we will \textbf{incorporate culturally-aware variants} trained on literary traditions beyond Western narratives. This requires collaboration with cultural experts to ensure authentic representation of story structures and conventions across global contexts~\citep{niu2001cultural}. Finally, we will \textbf{develop expertise-sensitive interfaces} that dynamically adjust scaffolding based on user proficiency. The system will track selection speed, edit frequency, and coherence scores to provide exploration support for novices and streamlined tools for experts~\citep{salomon1991partners}, adapting to both user capability and task demands.

\section{Conclusion}

The successful implementation of \ac{bvsr} theory in \model demonstrates that psychological creativity frameworks can provide effective blueprints for computational creativity systems. By explicitly separating variation generation from selective retention and implementing structured diversity through specialized personas, we have created a creative partnership model that enhances human creativity while preserving creative agency. The expertise-dependent preferences observed in our study highlight the importance of adaptive design in creative AI, while the strategic persona utilization patterns suggest that humans can effectively collaborate with \ac{ai} ensembles when provided with appropriate interaction frameworks.

These results suggest that the next frontier of creative AI lies not solely in smarter models, but in principled design that respects the nuances of individual creative cognition. As \ac{ai} capabilities continue to evolve, theoretically-informed approaches will be vital in ensuring that AI remains a partner in the creative process---fostering systems that genuinely extend the reach of human imagination while safeguarding the fundamental human drive for expression and ownership.

\begin{acks}

We extend our gratitude to Zhen Chen for creating the beautiful illustrations that enhance this paper. We also thank Yujia Peng, Guangyuan Jiang, and Yizhou Wang for their valuable feedback and insightful suggestions throughout the development of this work. This work is supported in part by the National Science and Technology Innovation 2030 Major Program (2025ZD0219400), the National Natural Science Foundation of China (62376009, 62376031), the State Key Lab of General \ac{ai} at Peking University, the PKU-BingJi Joint Laboratory for Artificial Intelligence, the Wuhan Major Scientific and Technological Special Program (2025060902020304), the Hubei Embodied Intelligence Foundation Model Research and Development Program, and the National Comprehensive Experimental Base for Governance of Intelligent Society, Wuhan East Lake High-Tech Development Zone.
\end{acks}

\bibliography{reference_header,reference}
\bibliographystyle{ACM-Reference-Format}

\end{document}

%% file: config.tex
\usepackage{graphicx} \graphicspath{{figures/}}
\usepackage{amsmath,mathtools,amsthm,nicefrac}
\usepackage[capitalise,nameinlink]{cleveref}
\usepackage[skip=3pt,font=small]{subcaption}
\usepackage[skip=3pt,font=small]{caption}
\usepackage[printonlyused]{acronym}
\usepackage{enumerate,enumitem}
\usepackage{xspace}
\usepackage[misc]{ifsym}
\usepackage{verbatim,fancyvrb,listings}
\usepackage{booktabs,tabularx,colortbl,multirow,multicol,array,makecell,tabularray}
\usepackage{fontawesome5}
\usepackage[normalem]{ulem}
\usepackage{textcomp}

\makeatletter
\DeclareRobustCommand\onedot{\futurelet\@let@token\@onedot}
\def\@onedot{\ifx\@let@token.\else.\null\fi\xspace}
\def\eg{\emph{e.g}\onedot}

\def\vs{\emph{vs}\onedot}

\def\etal{\emph{et al}\onedot}
\makeatother

\crefname{algorithm}{Alg.}{Algs.}
\Crefname{algocf}{Algorithm}{Algorithms}
\crefname{section}{Sec.}{Secs.}
\Crefname{section}{Section}{Sections}
\crefname{table}{Tab.}{Tabs.}
\Crefname{table}{Table}{Tables}
\crefname{figure}{Fig.}{Figs.}
\Crefname{figure}{Figure}{Figures}
\crefname{equation}{Eq.}{Eqs.}
\Crefname{equation}{Equation}{Equations}
\crefname{appendix}{Appx.}{Appxs.}
\Crefname{appendix}{Appendix}{Appendices}

\definecolor{gblue}{HTML}{4285F4}
\definecolor{gred}{HTML}{DB4437}
\definecolor{ggreen}{HTML}{0F9D58}
\definecolor{customblue}{HTML}{8EA2C7} 

\newcommand{\model}{NarrativeLoom\xspace}

\newcommand{\BV}{\textit{Blind Variation}\xspace}

\newcommand{\SR}{\textit{Selective Retention}\xspace}

\acrodef{bvsr}[BVSR]{Blind Variation and Selective Retention}
\acrodef{llm}[LLM]{Large Language Model}
\acrodef{lm}[LM]{Language Model}
\acrodef{ai}[AI]{Artificial Intelligence}
\acrodef{hci}[HCI]{Human-Computer Interaction}
\acrodef{rag}[RAG]{Retrieval-Augmented Generation }
\acrodef{ttcw}[TTCW]{Torrance Test for Creative Writing}